\documentclass[aps,prr,a4paper,floats,fleqn,footinbib,lengthcheck,twocolumn,tightenlines,balancelastpage,10pt,showpacs,superscriptaddress,notitlepage]{revtex4-1}
\usepackage{comment}
\usepackage{subfiles}
\usepackage{xr}
\usepackage{natbib}
\usepackage{amssymb}
\usepackage{amsfonts}
\usepackage{amsmath}
\usepackage{mathtools}
\usepackage{graphicx}
\usepackage[usenames,dvipsnames]{color}
\usepackage{longtable}
\usepackage{revsymb}
\usepackage{units}
\usepackage{nicefrac}
\usepackage{textcomp}
\usepackage{epsfig}
\usepackage{footmisc}
\usepackage{soul}
\usepackage[version=4]{mhchem}
\usepackage{lipsum}    
\usepackage[overload]{textcase}
\usepackage{titlesec}
\providecommand{\U}[1]{\protect\rule{.1in}{.1in}}

\ifx\pdfoutput\relax\let\pdfoutput=\undefined\fi
\newcount\msipdfoutput
\ifx\pdfoutput\undefined\else
\ifcase\pdfoutput\else
\ifx\paperwidth\undefined\else
\ifdim\paperheight=0pt\relax\else\pdfpageheight\paperheight\fi
\ifdim\paperwidth=0pt\relax\else\pdfpagewidth\paperwidth\fi
\fi\fi\fi
\begin{document}

\title{Defect-engineered, universal kinematic correlations between superconductivity and Fermi liquid transport}
\author{M. ElMassalami}
\affiliation{Instituto de F\'{\i}sica, Universidade Federal do Rio de Janeiro, Caixa Postal 68528, 21941-972 Rio de Janeiro RJ, Brazil}
\author{M. B. Silva Neto}
\affiliation{Instituto de F\'{\i}sica, Universidade Federal do Rio de Janeiro, Caixa Postal 68528, 21941-972 Rio de Janeiro RJ, Brazil}

\begin{abstract}

Identifying universal scaling relations between two or more variables in a complex system  
has always played a pivotal role in the understanding of various phenomena in different branches 
of science. Examples include the allometric scaling among food webs in biology, the scaling 
relationship between fluid flow and fracture stiffness in geophysics, and the so called 
gap-to-$T_c$ ratio, between the quasi-particle gap, $\Delta$, and the transition 
temperature, $T_c$, hallmarks of superconductivity. Kinematics, in turn, is the branch of 
physics that governs the motion of bodies by imposing constraints correlating their masses, 
momenta, and energy; it is therefore an essential ingredient to the analysis of e. g. high-energy 
quarkonium production, galaxy formation, and to the $\rho_\circ+AT^2$ contribution to the normal state 
resistivity in a Fermi liquid (FL), with $\rho_\circ$ being a measure of disorder and $A$ the 
hallmark of FL. Here, we report on the identification of a novel, universal kinematic scaling 
relation between $T_c(\rho_\circ)$ and $A(\rho_\circ)$, empirically observed to occur in a 
plethora of defect-bearing, conventional, weakly- or strongly- coupled superconductors, within 
their FL regimes. We traced back this relation to the triggering and stabilization of an
indirect electron-electron scattering channel, inside a very specific, yet common, type of 
amorphized region, ubiquitous in all such superconductors. Our theoretical formulation of the
problem starts with the construct of a distorted lattice, as a mimic of the kinematic aftermath 
to the formation of such amorphized regions. Then we apply standard many body techniques to derive 
analytic expressions for both $T_c(\rho_\circ)$, by using Eliashberg's theory of superconductivity, 
and for $A(\rho_\circ)$, from Boltzmann's quantum transport theory, as well as their 
mutual correlations. Our results are in remarkable agreement with experiments and provide 
a solid theoretical foundation for reconciling superconductivity with FL transport in these systems.
\end{abstract}
\maketitle

\section{Introduction}

\begin{figure*}[ht]
\begin{centering}
\includegraphics[scale=0.29]{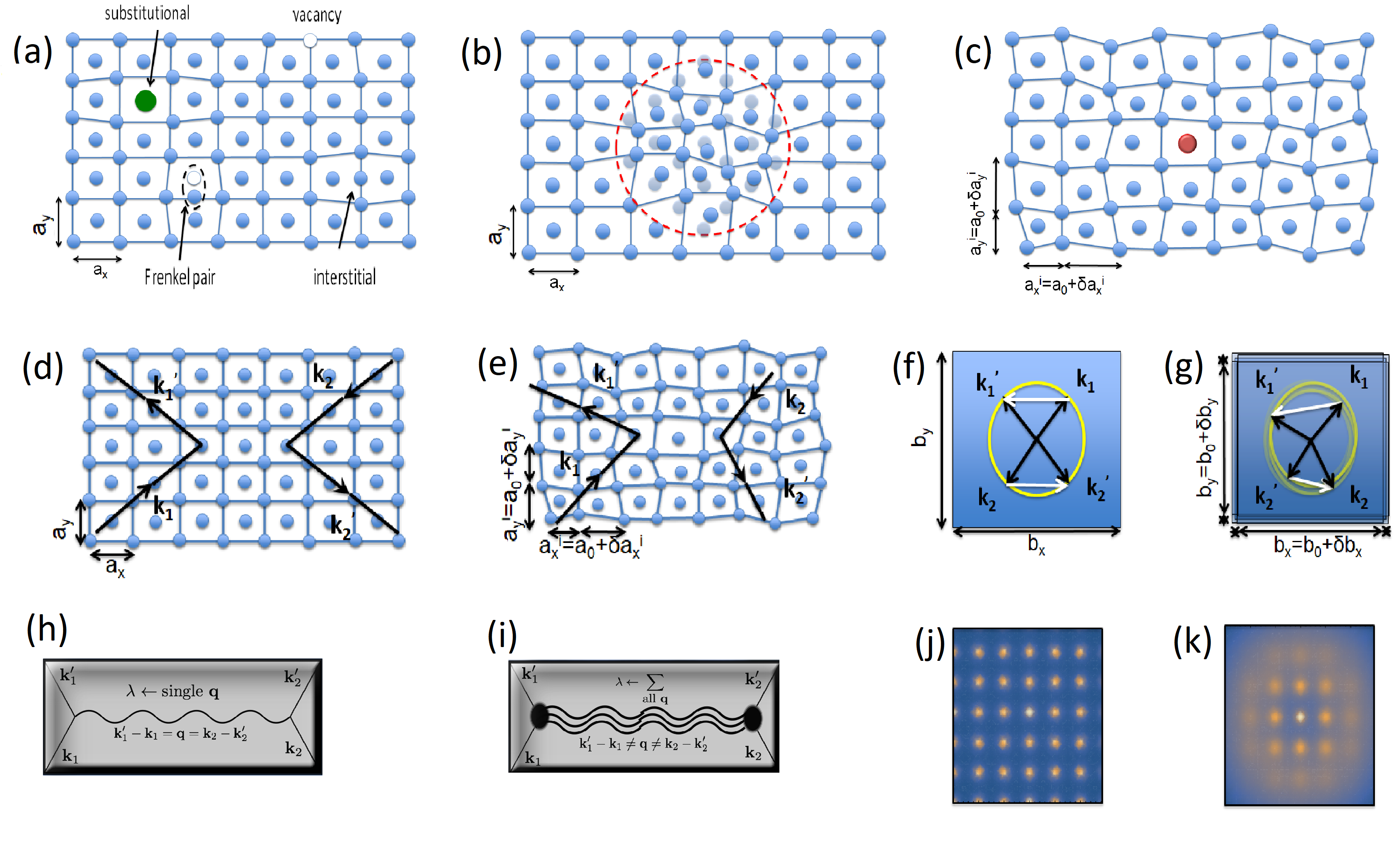}
\par\end{centering}
\caption{(a) Point defects in a crystalline solid; (b) agglomerated defects inside a spherical
region, limited by a red dashed radius, $nm \ll D \ll  \mu m$, and labelled as a defectal;  
(c) model of a defectal in terms of a distorted lattice plus a heavy scatterer. Effective 
two-electron scattering process occurring on the lattice, (d-e), in the first Brillouin zone,
(f-g), and in terms of Feynman diagrams, (h-i). The Fraunhoffer diffraction pattern for 
pristine (j) and distorted (k) structures, in which a typical amorphization halo is observed 
at higher-order reciprocal lattice points.  
}
\label{Fig1-Point-Defectals-Model}
\end{figure*}

Real crystalline solids do not manifest perfect atomic arrangement; rather, some degree of 
imperfection is always present in the form of naturally-occurring or artificially-engineered 
defects \cite{Ehrhart91-Defects-Metals-Alloys}. The various types of defects, e.g., point, linear, 
planar or bulk, Fig.~\ref{Fig1-Point-Defectals-Model}(a), can be intentionally 
engineered via techniques such as co-deposition, ion implantation, irradiation, 
chemical substitution, or thermal treatment. Manipulation of the type, concentration and distribution of these 
defects can lead to dramatic variations in the mechanical, thermal, optical, and electronic properties
of the host matrix. Such powerful leverage has been extensively employed by both academics and 
applied scientists for engineering highly-desirable technological marvels such 
as, stainless steel, semiconductor electronic components and high-temperature superconductors.

Remarkable as it is, defect engineering introduces additional features that have neither been explored 
nor are fully understood. Most of these features can be demonstrated by considering Aluminum thin films as a 
working example. Al films, less than $1 \mu$m thick and free of intentionally-incorporated defects, are characterized by small residual 
resistivities, $\rho_\circ \simeq 10 \,\mu\Omega$cm, low superconducting transition temperatures,
$T_c \simeq 1.2$ K, and normal-state resistivities largely dominated by the 
Bloch-Gruneisen, $T^5$, power law, typical of scattering between electrons and phonons,
with only a negligible electron-electron, $AT^2$, contribution (a small Fermi-liquid 
coefficient, $A \simeq 10^{-7}\mu\Omega$cm/K$^2$). 
Implantation/co-deposition of a few percents of oxygen into such Al film, {\it simultaneously}, gives rise to: 
(i) a huge increase in the residual resistivity $\rho_\circ$ ($10\, \mu\Omega$cm to $10^4 \mu\Omega$cm); 
(ii) an order of magnitude variation in $T_c$ ($1$ K to $10$ K)
\cite{Ziemann78-Al-granularFilms,Ziemann79-Al-Self-Irradiated-Films,Miehle92-SC-Implantation-Metals,Bachar13-Kondo-granular-Al,Bachar13-Kondo-granular-Al,17-Al-ThinFilms-O-irradiation}, and 
(iii) the surge and/or stabilization of a robust Fermi-liquid coefficient $A$ 
(from $10^{-7}\mu\Omega$cm/K$^2$ to $\mu\Omega$cm/K$^2$)
\cite{Ziemann79-Al-Self-Irradiated-Films}.  
These are impressive and unexpected features, to say the least, and constitute a long standing puzzle for many
reasons. First, oxygen is non-magnetic and Al is a conventional, isotropic (s-wave) superconductor. 
As dictated by Anderson's theorem
\cite{Anderson59-Theorem-disorder}, one would not expect changes in $T_c$; yet, $T_c$ is unambiguously enhanced. 
Second, although the Fermi surface of Al is large and disconnected, 
the phase space available for momentum relaxation is severely limited by kinematics and provides 
a negligibly small nominal value  for the Fermi-liquid coefficient $A$ 
\cite{MacDonald80-e-F-enhanced-e-e-Interaction}. As such, a $T^2$ contribution to the 
low temperature resistivity should only become relevant below 
$2$ K, right before superconductivity sets in \cite{MacDonald80-e-F-enhanced-e-e-Interaction}; 
yet, an overwhelmingly dominant Fermi-liquid contribution, 
manifested as large values for $A$, is triggered and stabilized by 
oxygen implantation/co-deposition over a wider temperature range. 
Finally, while $\rho_\circ$ is determined by the electron-impurity scattering, $\rho_\circ\sim|V_{imp}|^2$,
$T_c\sim e^{-1/\lambda}$ is associated with electron-phonon coupling, $\lambda \sim |V_{ep}|^2$, 
and $A$ to electron-electron interaction, $A\sim |V_{ee}|^2$. One would then expect
$\rho_\circ$, $T_c$ and $A$ to be independent;
yet, the variations of $T_c(\rho_\circ)$ and $A(\rho_\circ)$ are observed to be markedly
correlated as $\rho_\circ$ is increased by oxygen implantation/co-deposition. 
Most remarkably, we have found, after a thorough examination of the vast amount of 
available experimental data, scattered throughout 
the literature and collected over many decades, that the same three modifications in $\rho_\circ$, $T_c$, 
and $A$, as well as their mutual correlations, are manifested in several other conventional 
superconducting solids whenever defects are 
{\it properly engineered}, via different disordering techniques. 

In this work we show that all conventional, electron-phonon superconductors, 
both weakly- and strongly-coupled, when properly {\it defect-engineered}, 
exhibit a low temperature resistivity that can be universally written as
\begin{equation}
\rho(T)=\Theta(T-T_c)\left[\rho_\circ+A T^2\right],
\end{equation} 
with $\Theta(T-T_c)=1(0)$, for $T>(<)~~T_c$, where, in addition, 
$\rho_\circ,\, T_c$, and $A$ are uniquely correlated. 
We traced back the modifications observed in $\rho_\circ$, $T_c$, and $A$ to 
the same mechanism: the triggering and stabilization of an indirect 
electron-electron scattering channel, inside a very specific, yet common, 
type of amorphized region, ubiquitous in all such superconductors. We start 
with the construct of a distorted lattice, as to mimic the kinematic aftermath 
to the formation of such amorphized regions and we then apply standard many 
body techniques to derive analytic expressions for both $T_c(\rho_\circ)$, 
by using Eliashberg's theory of superconductivity, and for $A(\rho_\circ)$, 
from Boltzmann's quantum transport theory. Finally, we establish theoretically 
their mutual correlations and discuss our findings in connection to experiments.
%
%

\section{Defectals: description and implications\label{SubSec.Defectal-description}}

Analysis of the properties of defect-bearing samples reveals that only a certain 
type of stabilized agglomerate of defects \cite{Stritzker79-SC-Pd-Irradiated} is capable of 
producing those ubiquitous, though exotic, defect-related correlations between 
$\rho_\circ, T_c$, and $A$. For the identification of that specific type of defect agglomerate, see Fig.~\ref{Fig1-Point-Defectals-Model}(b), let us revisit our working example of Al film. 
Upon either implantation, irradiation, or 
co-deposition, large granular regions, containing oxygen-induced agglomerated 
disturbances, are observed in the host material.  Although
these three defect-incorporating 
processes are quite different in their experimental setups, and in the mechanism behind 
defects formation, distribution in size, and arrangement, they all manifest, nevertheless, similar 
influences on the normal and superconducting properties  of the target material, 
as it was empirically demonstrated 
for the case of oxygen implanted/co-deposited Al thin films \cite{17-Al-ThinFilms-O-irradiation}.
 
We envisage that an implanted/co-deposited oxygen acts as an active 
anchor that leads to the creation and/or stabilization of large amorphized disturbances. 
Each separate agglomerate can be thought of as a three-dimensional 
disordered metallic granule embedded in an otherwise perfectly arranged metallic host, Fig.~\ref{Fig1-Point-Defectals-Model}(b). 
Each of these 3-d agglomerate of defects within which an effective electron-electron scattering 
channel can be opened is labeled as a {\it defectal} and sketched in Fig.~\ref{Fig1-Point-Defectals-Model}(b).

Inasmuch the same way as with our working example of oxygen implanted/irradiated/co-deposited 
Al thin films, defectals can also be 
engineered in most simple metals or metallic alloys through
any disordering technique (e.g. quenched condensation, cold working, alloying, electron irradiation, etc) provided 
some sort of stabilizing anchor such O, H, etc, is present. 
Before entering into the details of our theoretical formulation for the 
consequences of defectal incorporation in conventional superconductors, 
let us revise some empirical curves depicting the evolution of $T_c(\rho_\circ)$ 
and $A(\rho_\circ)$ for a variety of systems.

\subsection{Correlation $T_c\times\rho_\circ$}

Figure \ref{Fig2-Exp-dTc-dRo-BCS} shows the evolution of 
$\frac{T_c - T^\circ_{c}}{T^\circ_{c}} ~ vs ~ \frac{\rho_\circ-\rho^\circ_\circ}{\rho^\circ_\circ}$ for a variety of defectal-incorporated 
materials. Here, $\rho^\circ_\circ$ and $T^\circ_c$ are the initial values of the residual resistivity and superconducting 
transition temperature, respectively, before the intentional addition of defectals. As such,
$\delta\rho_\circ\equiv\rho_\circ - \rho^\circ_\circ$ is a measure of the amount of the intentionally-introduced 
defectals.
In spite of the extensive list of differing materials and/or disordering techniques, the evolution of 
$\frac{\delta T_c}{T^\circ_{c}} ~ vs ~ \frac{\delta\rho_\circ}{\rho^\circ_\circ}$ can be classified into two distinct 
categories, discussed below. 

\subsubsection{Weakly-coupled superconductors} 

Figs.~\ref{Fig2-Exp-dTc-dRo-BCS}(b-c), shows the cases of In  \cite{Heim78-In-Implantation,Bergmann69-In-Sn-ee,Bauriedl76-In-Irradiation,Hofmann81-In-Irradiation,Miehle92-SC-Implantation-Metals}, Zn 
\cite{Miehle92-SC-Implantation-Metals}, Ga \cite{Miehle92-SC-Implantation-Metals,Gorlach82-Ga-Irradiation}, 
Al$_2$Au \cite{Miehle92-SC-Implantation-Metals}, AuIn$_2$ \cite{Miehle92-SC-Implantation-Metals}, Sn \cite{Bergmann69-In-Sn-ee}, and 
Al \cite{Miehle92-SC-Implantation-Metals,Ziemann79-Al-Self-Irradiated-Films}. 
Various other weakly-coupled superconductors (e.g. simple metals Be, Zn, Cd)
\cite{Ochmann83-Metal-H-SC-enhancement} can be added to this list. A common 
property is  that
defectal incorporation leads to an {\it enhancement} of $T_c$ and $\rho_\circ$, linear for small $\delta\rho_\circ$, 
$\delta T_c\propto\delta\rho_\circ$, with a slope that depends solely on material properties (see the thick red line in Figs.~\ref{Fig2-Exp-dTc-dRo-BCS}(b-c)).
It is worth noting that the manifestation of such correlation in self-ion irradiation of aluminum film, Fig.~\ref{Fig2-Exp-dTc-dRo-BCS}(c), indicates that defectal formation and stabilization does not depend 
on the chemical character of the bombarding projectiles (provided there is oxygen as a stabilizing anchor). 

\subsubsection{Strongly-coupled superconductors}

\begin{figure*}[hbtp]
\begin{centering}
\includegraphics[scale=0.5]{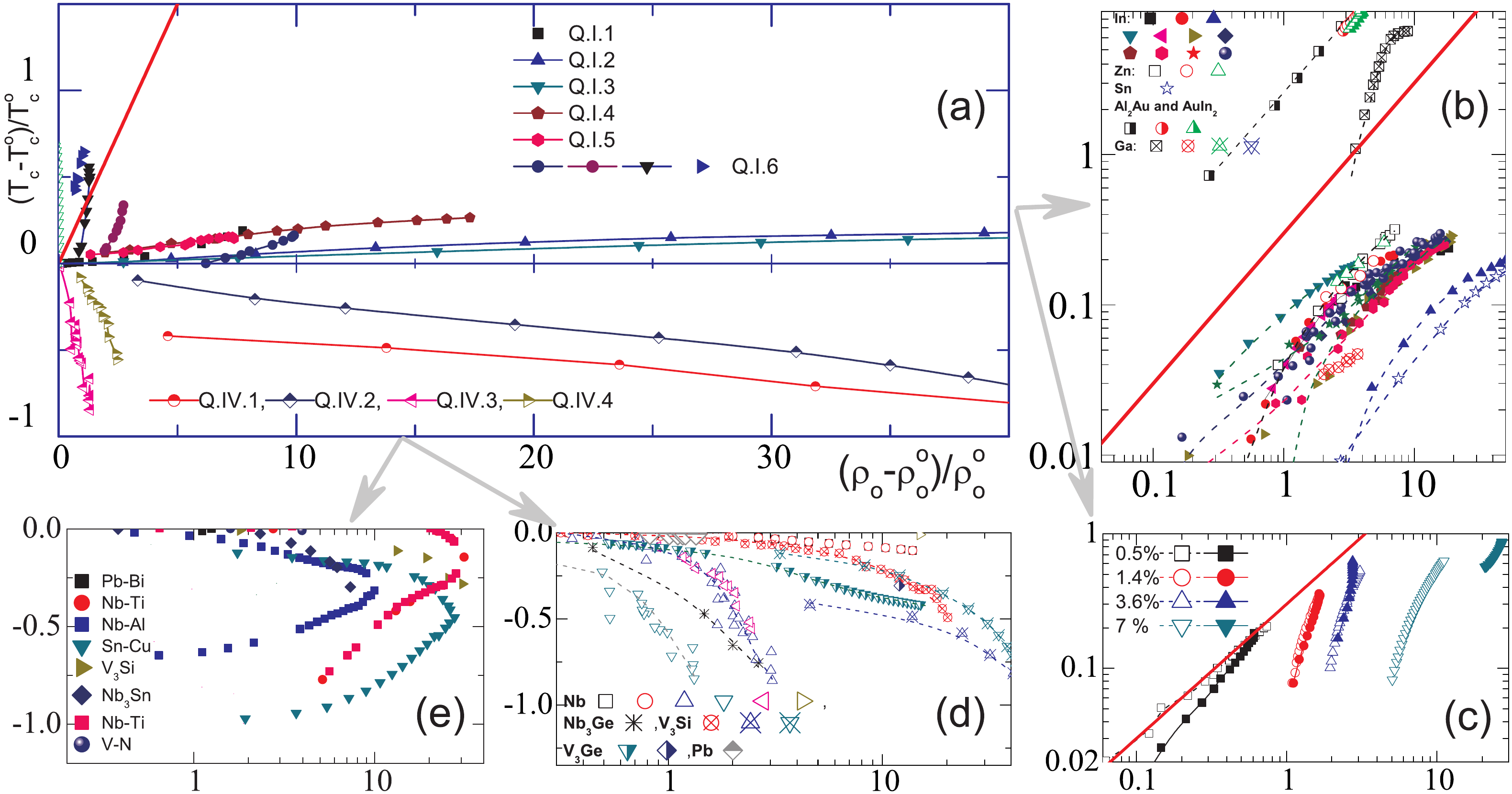}
\end{centering}
\caption{(a) Evolution of $\frac{\delta T_c}{T_c^\circ} \, ~ vs ~ \, \frac{\delta \rho_\circ }{\rho_\circ^\circ}$ for various defectal-bearing superconductors.  \underline{Weak-coupled}: Q.I.1: Al and O co-deposition  \cite{Ziemann78-Al-granularFilms}, Q.I.2: Indium-granular film \cite{Bergmann69-In-Sn-ee}, Q.I.3: Sn~-~granular \cite{Bergmann69-In-Sn-ee}, Q.I.4: In~-~Ar-irradiated \cite{Hofmann81-In-Irradiation}, Q.I.5: In~-~Ar-implanted \cite{Hofmann81-In-Irradiation}, Q.I.6: O-incorporated Al with Al-irradiation \cite{Ziemann79-Al-Self-Irradiated-Films}. 
\underline{Strong-coupled}: Q.IV.1: V$_{3}$Si Kr-irradiated \cite{Meyer80-Review-SC-ionImplantation},  Q.IV.2: V$_{3}$Si He-irradiated \cite{Meyer80-Review-SC-ionImplantation}, Q.IV.3: Nb N-implantated \cite{Linker80-Nb-Irradiation}, Q.IV.4: Nb Ne-irradiated \cite{Linker80-Nb-Irradiation}. 
(b) Log-log plots of irradiated/implanted thin-films of weakly-coupled conventional BCS superconductors: In \cite{Heim78-In-Implantation,Bergmann69-In-Sn-ee,Bauriedl76-In-Irradiation,Hofmann81-In-Irradiation,Miehle92-SC-Implantation-Metals}, Zn \cite{Miehle92-SC-Implantation-Metals}, Ga \cite{Miehle92-SC-Implantation-Metals,Gorlach82-Ga-Irradiation}, \ce{Al2Au} \cite{Miehle92-SC-Implantation-Metals}, \ce{AuIn2}  \cite{Miehle92-SC-Implantation-Metals}, and Sn \cite{Bergmann69-In-Sn-ee}.
(c) log-log plots of self-ion irradiation of pure (empty symbols) and granular (filled symbols) 
thin-films of aluminum with varying concentration of defectal-stablizing oxygen ($\leq 7 \%$) \cite{Miehle92-SC-Implantation-Metals,Ziemann79-Al-Self-Irradiated-Films}. 
(d) Semi-log plots of the strong-coupled conventional BCS superconductors: Nb \cite{Heim75-Nb-Implantation,Linker80-Nb-Irradiation}, \ce{Nb3Ge} \cite{Testardi77-A15-Resistvity},  \ce{V3Si} \cite{Testardi78-A15-LifeTime,Meyer80-Review-SC-ionImplantation},  \ce{V3Ge} \cite{Testardi77-A15-Resistvity}, \ce{Pb} \cite{Miehle92-SC-Implantation-Metals}, and \ce{Pb$_{1-x}$Ge$_x$} (x=0.3,0.7) \cite{Zawislak81-SUC-PbGe}. 
(e) Semi-log plot of irradiated/implanted thin-films of strong-coupled conventional BCS superconductors: \cite{Gurvitch86-Disorder-induced-transition-AT2}.
For more details, see text.
}
\label{Fig2-Exp-dTc-dRo-BCS}
\end{figure*}

Fig.~\ref{Fig2-Exp-dTc-dRo-BCS}(h-i),
shows the cases of V$_{3}$Si \cite{Meyer80-Review-SC-ionImplantation,Meyer80-Review-SC-ionImplantation}, 
Nb \cite{Linker80-Nb-Irradiation,Heim75-Nb-Implantation,Linker80-Nb-Irradiation}, 
\ce{Nb3Ge} \cite{Testardi77-A15-Resistvity},  \ce{V3Si} \cite{Testardi78-A15-LifeTime,Meyer80-Review-SC-ionImplantation},  
\ce{V3Ge} \cite{Testardi77-A15-Resistvity}, \ce{Pb} \cite{Miehle92-SC-Implantation-Metals}, and 
\ce{Pb$_{1-x}$Ge$_x$} (x=0.3, 0.7) \cite{Zawislak81-SUC-PbGe}. 
In this class, defectal incorporation leads to a {\it reduction in $T_c$} together with an increase in 
$\rho_\circ$. Although Pd is a non-superconducting metal in the pure state, its 
hyrdogenation \cite{Stritzker75-Pd-Implantation,Stritzker78-Pd-SC-disorder,Stritzker79-SC-Pd-Irradiated}, 
or that of its solid-solution Pd-\textit{X} (\textit{X}=noble metal) \cite{Stritzker74-SUC-Pd-NobleMetals}, 
leads to a superconducting state with a relatively high $T_c$. A common property of this class is the {\it continuous nonlinear dependence} of $\delta T_c$ on $\delta\rho_\circ$, for 
all $\delta\rho_\circ\geq 0$, and a negative derivative, $d(\delta T_c)/d(\delta\rho_\circ)<0$, for a 
majority of its members. Finally, the parabolic-like behaviour observed in Fig.~\ref{Fig2-Exp-dTc-dRo-BCS}(h) 
is a characteristic feature of a superconducting binary alloy, e.g. $A_{1-x}B_{x}$, wherein 
the residual resistivity is non-monotonic and follows Nordheim's rule $\rho_\circ(x) \propto x(1-x)$.

\subsection{Correlation $A\times\rho_\circ$}

Figure~\ref{Fig3-Exp-Tc-A-Ro-Exp-lnTc-sqrA} shows the 
defectal-induced evolution of $T_c$, $A$ and $\rho_\circ$ for selected representatives from each of the 
aforementioned two classes: the incorporation of defectals 
leads to the surge of $AT^2$ Fermi-liquid contribution, with $A$ being strongly correlated 
to $\rho_\circ$. It is remarkable that Gurvitch  \cite{Gurvitch86-Disorder-induced-transition-AT2} 
had already identified the importance of disorder-driven breakdown of momentum conservation 
in shaping $T_c$, $A$ and $\rho_\circ$ of superconducting alloys. Unfortunately, with the exception of 
that work \cite{Gurvitch86-Disorder-induced-transition-AT2}, such a correlation had not been highly 
appreciated. As such, there are no extensive reports from which one can construct a universal 
$\frac{A - A_\circ}{A_\circ} ~vs~ \frac{\rho_\circ-\rho^\circ_\circ}{\rho^\circ_\circ}$ plot. Nevertheless, a common, 
parabolic-like dependence of $A$ on $\rho_\circ$, like $A(\rho_\circ)=A_\circ + A_1 \rho_\circ + A_2 \rho_\circ^2$,
can be readily identified when examining the evolution of $A$ in the 
representatives of: (i) the strongly-coupled [Fig.~\ref{Fig3-Exp-Tc-A-Ro-Exp-lnTc-sqrA}(a.8)]; and 
(ii) the weakly-coupled [Fig.~\ref{Fig3-Exp-Tc-A-Ro-Exp-lnTc-sqrA}(b.1)] superconductors. 

\subsection{Correlation $T_c\times A$} 

\begin{figure*}[hbtp]
\centering
\includegraphics[scale=0.5]{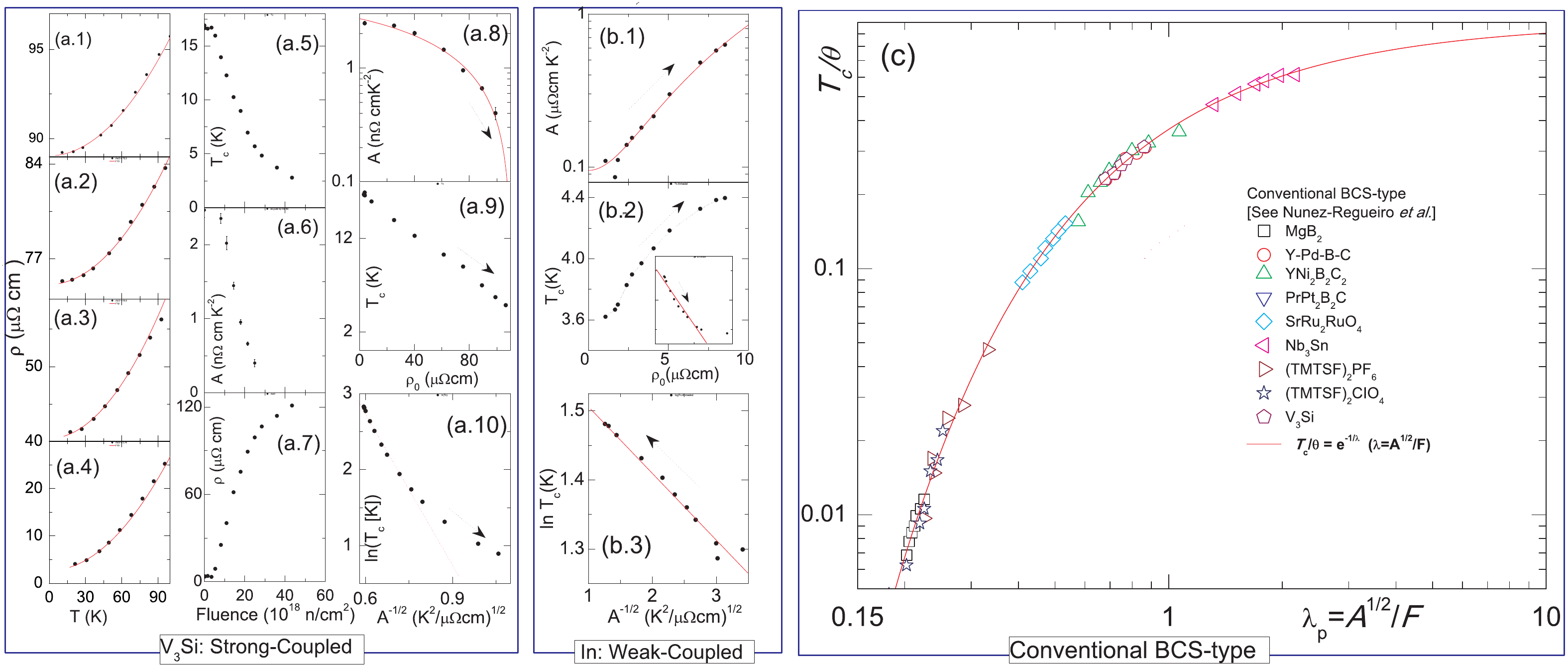}
\caption{(a) Bulk single-crystal \ce{V3Si} subjected to irradiation by neutron flux [data taken from Ref.\onlinecite{Caton82-A15-Irradiated}]. 
$\rho_\circ (T)$ with fluences  (a.1) $21.5\times 10^{18} n/cm^{2}$,  (a.2) $18\times 10^{18} n/cm^{2}$,  (a.3) $11\times 10^{18} n/cm^{2}$, and  (a.4) $0\times 10^{18} n/cm^{2}$. The red lines are fit to $\rho _{tot}(T) =\rho_\circ+\beta T^{5}+AT^{2}$. (a.5) $T_c$ \textit{~ vs ~} fluence, (a.6) $A$ \textit{~ vs ~} fluence,  and (a.7) $\rho_\circ$ \textit{~ vs ~} fluence. (a.8) $A$ \textit{~ vs ~} $\rho_\circ$, (a.9) $T_c$ \textit{~ vs ~} $\rho_\circ$, and (a.10) $ln(T_c)$ \textit{~ vs ~} $A^{-\frac{1}{2}}$; 
(b) thin films of In deposited at 200K and annealed at 300K; afterwards implantated with In$^{+}$ ions at 2K [data taken from Ref.\onlinecite{Heim78-In-Implantation}]: (b.1) $A$ \textit{~ vs ~} $\rho_\circ$,  (b.2) $T_c$ \textit{~ vs ~} $\rho_\circ$, \textit{Inset}: $lnT_c$ \textit{~ vs ~} $\rho_{\text{o}}^{\text{-1}}$,  and (b.3) $ln(T_c)$ \textit{~ vs ~} $A^{-\frac{1}{2}}$. 
(c) Universal kinematic correlation between $T_c$ and $A$ for several weakly- and strongly- coupled BCS superconductors [data taken from Ref.\onlinecite{Nunes12-FermiLiquid-SUC}].
}\label{Fig3-Exp-Tc-A-Ro-Exp-lnTc-sqrA}
\end{figure*}

Figure \ref{Fig3-Exp-Tc-A-Ro-Exp-lnTc-sqrA} also reveals a remarkable universal correlation 
between $T_c$ and $A$, with a BCS-like form, $T_c=\theta e^{-F/\sqrt{A}}$, wherein $\theta$ and $F$ 
are material-dependent parameters, specific for each superconductor. This remarkable correlation has 
previously been recognized and theoretically approached in a few material systems (see, 
e.g., the seminal, pioneering  works of 
refs.~\cite{Gurvitch86-Disorder-induced-transition-AT2,Nunes12-FermiLiquid-SUC,vanderMarel11-SrTiO3-FL-SUC,18-Castro-Tc-A-Correlation}), but here we show that it can be unambigously traced back to the presence of defectals, being the main factor 
behind the establishment of such universal kinematic correlation between $T_c(\rho_\circ)$ and $A(\rho_\circ)$, 
and allowing us to construct a single $T_c ~vs~ \sqrt{A}$ plot [Figs.~\ref{Fig3-Exp-Tc-A-Ro-Exp-lnTc-sqrA}(a.10,b.3,c.6 and d.6) 
and ~\ref{Fig3-Exp-Tc-A-Ro-Exp-lnTc-sqrA}] that includes many representatives of each of the two classes described above. 
%

\section{The mechanism: distortion and softening\label{Sec.Theoretical-Analysis}}

Generally speaking, defectal incorporation is expected to lead to significant changes to the host 
lattice. Figure~\ref{Fig1-Point-Defectals-Model}(b) represents the authors' impression of a defectal, which 
shall henceforth be modeled in terms of {\it a distorted lattice plus a heavy scatterer}, see Fig.~\ref{Fig1-Point-Defectals-Model}(c). 
Albeit its simplicity, this model embodies the two most important outcomes following 
defectal incorporation: distortion and softening. The breakdown of translational invariance, 
and the consequent triggering of multiply-polarized, phonon-mediated, electron-electron 
interaction channels, as illustrated in 
Figs.~\ref{Fig1-Point-Defectals-Model}(e,~g,~i),
follows from distortion. In addition, the softening of the 
vibrational spectrum, and the consequent transfer of spectral weight to lower frequencies, follows from the 
development of low energy resonances at the lower edge of the spectrum, or quasi-localized phonon modes 
\cite{Lifshitz66-Dynamic-Defective-Crystals,Gorkov17-Mechanism-SrTiO3}, that originate from the scattering of phonon waves off heavy scatterers. As we shall soon demonstrate, these effects will determine the evolution 
of both superconductivity and normal-state transport.

\subsection{\bf Distortion - Hosemann's paracrystal} 

Let us first introduce the concept of a {\it distorted lattice}, or a Hosemann's paracrystal \cite{Hosemann}. 
Consider, for simplicity, the defectal-free structure to be a cubic lattice with primitive unit-cell 
vectors $|{\bf a}_{i=1,2,3}|=a_\circ$. We model the defectal-bearing structure as a distorted 
lattice wherein the "unit vectors" acquire a given statistical probability described by a Gaussian 
distribution in which the average $|\overline{\bf a}_{i=1,2,3}|=a_\circ$ corresponds to the center 
of the distribution, while the extent of the distortion is given by the Gaussian width 
$\sigma_{ij}=\overline{\Delta {a}_j}/\overline{{a}_i}$ \cite{Hosemann}. For simplicity, we assume
equal variance, $\sigma_{ij}=\sigma\delta_{ij}$, leading to the structure depicted in 
[Fig.~\ref{Fig1-Point-Defectals-Model}(c)], whose "unit-cell" vectors vary in length 
and direction from cell to cell, that can nevertheless be still organized in rows and columns. 
For long range crystal order, Bragg reflections occur at reciprocal lattice points, 
${\bf g}(h,k,l)=h{\bf b}_1+k{\bf b}_2+l{\bf b}_3$, spanned in terms of primitive vectors 
${\bf b}_{i=1,2,3}$ satisfying ${\bf a}_i\cdot{\bf b}_j=2\pi\delta_{ij}$. However,
for defectal-bearing structures, the amplitude of "Bragg reflections" is ever decreasing, 
while the line-width $\delta{\bf g}(h,k,l)$ at ${\bf g}(h,k,l)$ is monotonically broadening \cite{Hosemann}.
Fig.~\ref{Fig1-Point-Defectals-Model}(k) shows the Frauhoffer broadening of diffraction 
patterns in distorted lattices and how these differ from those of a pristine crystal, 
Fig.~\ref{Fig1-Point-Defectals-Model}(j).

Consider now the fate
of quasi-momentum conservation during a scattering of an electron by a phonon: 
{\it in a distorted lattice, an electron, initially at a state ${\bf k}_1$ that goes into a final state
${\bf k}_1^\prime$ after being scattered by a phonon with wavevector ${\bf q}$, transfers an 
amount of quasi-momentum ${\bf k}_1^\prime-{\bf k}_1-{\bf q}= {\bf g}+\bf {\delta g}$}. Evidently, 
for a defectal-free system, $\bf{\delta g}=0$, quasi-momentum is conserved exactly and
${\bf q}={\bf k}_1^\prime-{\bf k}_1-{\bf g}$; in contrast, for a defectal-bearing system, $\bf {\delta g}\neq 0$,
quasi-momentum is no longer conserved in the sense that ${\bf q}={\bf k}_1^\prime-{\bf k}_1-{\bf g}-\delta{\bf g}$ 
becomes increasingly arbitrary, especially those with $\bf {\delta g}$ in higher Brillouin zones. 

The above statement can be made mathematically precise with the aid of the electron-phonon 
structure factor, $S_{\bf q}({\bf k}_1^\prime-{\bf k}_1)=|\varphi({\bf k}_1^\prime-{\bf k}_1-{\bf q})|^2$,
written in terms of
$\varphi({\bf k}_1^\prime-{\bf k}_1-{\bf q})=(1/\sqrt{N})
\sum_{{\bf r}}e^{i({\bf k}_1^\prime-{\bf k}_1-{\bf q})\cdot{\bf r}}$, 
the electron-phonon interaction phase \cite{Ziman79-Electrons-Phonons},
as defined for a lattice containing $N$ ions at ${\bf r}=n_1{\bf a}_{1}+n_2{\bf a}_{2}+n_3{\bf a}_{3}$.  
For a pristine lattice 
$S_{\bf q}({\bf k}_1^\prime-{\bf k}_1)=\sum_{{\bf g}}\delta_{{\bf k}_1^\prime-{\bf k}_1-{\bf q},{\bf g}}$,
for either normal, ${\bf g}=(0,0,0)$, or umklapp, ${\bf g} \neq (0,0,0)$, scatterings. As in a
perfect crystal, the low temperature ion deformations are smooth and of long wavelength (small ${\bf q}$), 
one usually needs to retain normal events only and, as a result, only longitudinal phonon modes with 
a well defined polarization, $\hat{{\bf e}}({\bf q}={\bf k}_1^\prime-{\bf k}_1)$, are excitable. 
In contrast, in a defectal-bearing system the electron-phonon structure factor acquires broadened 
features at ${\bf g}\neq 0$ because granular distortions provides a source of short wavelength (large ${\bf q}$) 
phase interference. Within the framework of the distorted lattice, these features are intrinsic, as one 
can verify by looking at the averaged structure factor calculated in appendix \ref{El-Ph-Coupling-Defectal}
\begin{equation}
\overline{S}^\ell_{\bf q}({\bf k}_1^\prime-{\bf k}_1)
\approx \delta_{{\bf k}_1^\prime-{\bf k}_1-{\bf q},0}
+\sum_{{\bf g}\neq 0}\frac{S_{max}({\bf g})}{1+\ell^2({\bf k}_1^\prime-{\bf k}_1-{\bf q}-{\bf g})^2}.
\label{disordered-structure-factor-2}
\end{equation}
In the above expression, the peak amplitudes are 
$S_{max}({\bf g})=4/\sigma^2 {\bf g}^2=a_\circ^2/\sigma^2\pi^2(h^2+k^2+l^2)$, while 
$\ell\equiv|\delta{\bf g}|^{-1}=4/\sigma^2{\bf g}^2a_\circ=a_\circ/\sigma^2\pi^2(h^2+k^2+l^2)$, 
is inversely proportional to the widths of its peaks, $\delta{\bf g}$ \cite{Hosemann}, see 
appendix \ref{El-Ph-Coupling-Defectal}. Since now 
${\bf k}_1^\prime-{\bf k}_1 - {\bf q}-{\bf g}\neq 0$, multiple phonon modes 
(longitudinal and transverse, of all polarizations 
$\hat{{\bf e}}({\bf q}\neq {\bf k}_1^\prime-{\bf k}_1-{\bf g})$) become kinematically 
available for mediating electron-electron interactions.  

Let us consider the parameter $\ell$ as an effective "mean-free path" which 
is proportional to the inverse of $\delta \rho_\circ$ (the increase in the residual resistivity 
due to scattering of electrons off the defectals)
\begin{equation}
\frac{\delta\rho_\circ}{\rho_\circ^\circ}=\frac{\ell_\circ}{\ell},\;\mbox{with}
\;\ell_\circ=a_\circ\left(\frac{\rho_\circ^{am}}{\rho_\circ^\circ}-1\right),
\;\mbox{for}\quad\rho_\circ^{am}\gg\rho_\circ^\circ,
\label{Eq-defining-L}
\end{equation}
where $\rho_\circ^\circ$ and $\ell_\circ$ are the initial values, while $\rho_\circ^{am}$
is the residual resistivity for the amorphous case. Effectively, $\ell$ is a 
scaling length related to the degree of distortion in the primitive unit-cell vectors ($\sigma$) and, 
as such, can be used to continuously interpolate between two limits: the defectal-free 
case crystal ($\ell\rightarrow\infty$, $\rho_\circ\rightarrow \rho_\circ^{\circ}$) and the 
amorphous, neighboring defectals, regime ($\ell\rightarrow a_\circ$, $\rho_\circ\rightarrow \rho_\circ^{am}$).

\subsection{\bf Softening - Lifshitz's resonance} 

Next we elaborate on the notion of a {\it heavy scatterer}, or a Lifshitz's
resonance \cite{Lifshitz66-Dynamic-Defective-Crystals}. Each one 
of such large collection of misplaced and/or implanted atoms, precipitates and/or granular 
amorphous phases, being part of a defectal or as individual entities, can also be seen, from 
the point of view of long wavelength phonon waves, as a heavy scatterer. This leads to a 
slowing down of long wavelength vibrations and to an important transfer of spectral 
weight towards the lower edge of the spectrum. 
If the amplitude of the incident and scattered phonon waves are 
$\varphi_{{\bf q},\nu}^{(i)}$ and $\varphi_{{\bf q},\nu}^{(s)}$
\cite{Lifshitz66-Dynamic-Defective-Crystals}, respectively, then 
these two quantities are related by
\begin{equation}
\varphi_{{\bf q},\nu}^{(s)}=\frac{1}{1-\varepsilon D(\omega)}\varphi_{{\bf q},\nu}^{(i)},
\end{equation}
where $\varepsilon=({\cal M}-M)/M$, with ${\cal M}$ being an effective {\it defect-related mass}, and 
$D(\omega)$ is a function of only the frequency $\omega$. If the frequency of the driving wave 
lies inside the continuum of vibrations, especially at the bottom of the phonon bands, then 
$D(\omega)={\cal R}e[D(\omega)]+i{\cal I}m[D(\omega)]$,
and a resonance is found at a frequency $\omega_R$ given by the condition
$\varepsilon {\cal R}e[D(\omega_R)]=1$.
Generally, the function ${\cal R}e[D(\omega)]\sim\omega^2$ and the effective mass 
${\cal M}$ is much heavier than the typical mass of the lattice ion, $M$. 
Then, for sufficiently large $\varepsilon\gg 1$, the resonance frequency ($\omega_R\sim 1/\sqrt{\varepsilon}$) will be located at the low-frequency range of the spectrum \cite{Lifshitz66-Dynamic-Defective-Crystals}.

The phase shift due to the phonon-wave scattering off defectals can be written as
\cite{Lifshitz66-Dynamic-Defective-Crystals}
\begin{equation}
\Phi(\omega)=\arctan{\left[\frac{\varepsilon{\cal I}m[D(\omega)]}{1-\varepsilon {\cal R}e[D(\omega)]}\right]},
\end{equation}
which, when close to $\omega_R$, changes rapidly
from $0$ to $\pi$, indicating that the effective impurity oscillates out of phase with respect
to the underlying lattice ions. This acts as a driving force that produces 
the sharp resonance peak in the vibrational density of states ${\cal F}(\omega)$.
For a concentration $n_d$ of defectals this peak is given by 
\begin{equation}
\delta{\cal F}_R(\omega)=\frac{3}{\pi}\frac{d\Phi}{d\omega}\approx \frac{n_d}{2\pi}\frac{\Gamma}{(\omega-\omega_R)^2+\frac{1}{4}\Gamma^2},
\label{Eq.Spectral-Function-defectal-disordered-Resonance}
\end{equation}
and the width $\Gamma$ of the resonance at $\omega_R$ is given in terms of the phase shift, $\Phi(\omega)$, during 
the scattering of phonon waves off a dilute concentration, $n_d$, as 
\begin{equation}
\Gamma=\frac{2\pi {\cal F}(\omega_R)}{\left\{ d\Phi(\omega)/d\omega\right\}_{\omega_R}}.
\end{equation}
As we can see, the larger the mass ${\cal M}$, the lower the frequency 
$\omega_R$, since $\omega_R\sim 1/\sqrt{\varepsilon}$, and the sharper the resonance 
will be, as $\Gamma$ is proportional to ${\cal F}(\omega_R)$ \cite{Lifshitz66-Dynamic-Defective-Crystals}.

\subsection{Combining distortions and softening}

The combination of a broadened structure factor in the electron-phonon coupling and the existence of 
quasi-localized phonon modes lead to a generalized form for Eliashberg's spectral function
which has been calculated in appendix \ref{Eliashberg-Spectral-Function-Defectal}
\begin{widetext}
\begin{equation}
\alpha^{2}\mathcal{F}_\ell\left(\omega\right)=
\sum_{\left\{\mathbf{k}^{\prime},\mathbf{k}\right\}={\bf k}_F,\mathbf{q},\nu}
\overline{S}^\ell_{\mathbf{q}}\left(\mathbf{k}^{\prime}-\mathbf{k}\right)\left|g_{\mathbf{k}^{\prime},\mathbf{k},\mathbf{q},\nu}\right|^{2}
\left\{
\delta\left(\omega-\omega_{\mathbf{q},\nu}\right)+n_d(\ell)\frac{2}{\pi}\frac{\Gamma}{4(\omega-\omega_R)^2+\Gamma^2}\right\},
\label{Eq-Eliashberg-Spect-Function}
\end{equation}
\end{widetext}
where $g_{\mathbf{k}^{\prime},\mathbf{k},\mathbf{q},\nu}=\alpha(\omega_{{\bf q},\nu})
\hat{{\bf e}}({\bf q},\nu)\cdot \left({\bf k}^\prime-{\bf k}\right)$ is the amplitude of the electron-phonon matrix
element [including the bare $\alpha(\omega_{{\bf q},\nu})$ due to all branches, $\nu=L,T_1,T_2$, 
with dispersion $\omega_{\mathbf{q},\nu}$ and
polarization $\hat{{\bf e}}({\bf q},\nu)$, see appendix \ref{El-Ph-Coupling-Defectal}], while
$\omega_R$ and $\Gamma$ are, respectively, the frequency and linewidth of the low-energy, 
quasi-localized phonon resonances associated with a density, $n_d(\ell)$, of Lifshitz heavy 
scatters.

$\alpha^{2}\mathcal{F}_\ell\left(\omega\right)$ of Eq.(\ref{Eq-Eliashberg-Spect-Function}) summarizes, 
mathematically, our simple (distorted-lattice-plus-heavy-scatterer) defectal-model, 
as it includes: (i) the softening of the vibrational spectrum, through the continuous transfer of 
spectral weight, tracked by $\ell$, from high, Debye's, to low, $\omega_R$, frequencies; 
(ii) the inclusion of new phonon branches, $\nu$, and polarizations, $\hat{{\bf e}}({\bf q},\nu)$,
through $\sum_{{\bf q}, \nu}$; (iii) the sum of all kinematically unconstrained wave-vectors, 
$\mathbf{k}^{\prime},\mathbf{k},\mathbf{g},\mathbf{q}$ ($0\leq |{\bf q}| \leq 2k_F$) whose 
rules of momentum transfer are controlled by $\overline{S}^\ell_{\mathbf{q}}\left(\mathbf{k}^{\prime}-\mathbf{k}\right)$. 
We calculated $\alpha^{2}\mathcal{F}_\ell\left(\omega\right)$ of Eq.(\ref{Eq-Eliashberg-Spect-Function}) within 
the Debye model for phonons interacting with nearly-free electrons, 
see appendix \ref{Eliashberg-Spectral-Function-Defectal}: its evolution for different values of $\ell$, is 
shown in  Fig.~\ref{Fig5-FL-Mechanisms-Eff-Vee}(g).

\section{Superconductivity and FL transport}

After incorporating distortion and softening, let us consider the two-particle process, shown in 
Figs.~\ref{Fig1-Point-Defectals-Model}(e, g, i), wherein electrons, 
initially at states ${\bf k}_1$ and ${\bf k}_2$, scatter into final states ${\bf k}_1^\prime$ and
${\bf k}_2^\prime$ by the exchange of all kinematically unconstrained phonon modes ${\bf q}$ with a nonzero spectral weight. 
The resulting retarded, attractive electron-electron interaction reads
\begin{equation}
V_{{\bf k}_1^\prime,{\bf k}_1,{\bf k}_2^\prime,{\bf k}_2}({\bf q},\ell) 
\approx -\phi_{{\bf q}}({\bf k}_1^\prime,{\bf k}_1,{\bf k}_2^\prime,{\bf k}_2)V_{ee}(\ell),
\label{Eq-Veff}
\end{equation}
and fully expounds the roles of distortions and softening through its phase,
$\phi_{{\bf q}}({\bf k}_1^\prime,{\bf k}_1,{\bf k}_2^\prime,{\bf k}_2)=(1/N)
\sum_{{\bf r}}e^{i({\bf k}_{1}^\prime-{\bf k}_1-{\bf q})\cdot{\bf r}}\sum_{{\bf r}^\prime}e^{i({\bf k}_{2}^\prime-{\bf k}_2+{\bf q})\cdot{\bf r}^\prime}$,
and amplitude, $V_{ee}(\ell)$, obtained after averaging over Fermi and Debye surfaces.
Now let us consider the influence of distortions and softening on $T_c(\ell)$, $A(\ell)$,
and their correlations. 

\subsection{$T_c(\ell)$ from Eliashberg's theory} 

$T_c(\ell)$ will be calculated as the zero gap limit, 
$\Delta_\ell\rightarrow 0$, of the system of Eliashberg's equations in imaginary time ($i\omega_n=i\pi T(2n-1)$)
\begin{widetext}
\begin{eqnarray}
\Delta_\ell(i\omega_n)Z_\ell(i\omega_n)&=&
\pi T\sum_m\left[\lambda_\ell(i\omega_m-i\omega_n)-\mu^*(\omega_c)\theta(\omega_c-|\omega_m|)\right]
\frac{\Delta_\ell(i\omega_m)}{\sqrt{\omega_m^2+\Delta_\ell^2(i\omega_m)}},\nonumber\\
Z_\ell(i\omega_n)&=&1+\frac{\pi T}{\omega_n}\sum_m 
\lambda_\ell(i\omega_m-i\omega_n)\frac{\omega_m}{\sqrt{\omega_m^2+\Delta_\ell^2(i\omega_m)}},
\end{eqnarray}
\end{widetext}
where the Coulomb pseudo-potential  
\begin{equation}
\mu^*(\omega_c)=\frac{\mu}{1+\mu\ln{\left(\frac{\epsilon_F}{\omega_c}\right)}},
\end{equation}
is given in terms of the bare, repulsive Coulomb interaction, $\mu=N(\epsilon_F)V_C$, and 
the cutoff frequency $\omega_c$, while $\lambda_\ell$ is the distortion-influenced electron phonon coupling
\begin{eqnarray}
\lambda_\ell(i\omega_m-i\omega_n)&=&
2\int_0^\infty \frac{\omega\; \alpha^2{\cal{F}_\ell}(\omega)}{(\omega_m-\omega_n)^2+\omega^2}d\omega\nonumber\\
&\equiv& N(\epsilon_F)V_{ee}(i\omega_m-i\omega_n;\ell),
\end{eqnarray}
wherein  $V_{ee}(i\omega_m-i\omega_n;\ell)$ is the retarded, attractive, effective electron-electron 
interaction in a distorted lattice, mediated by phonons. As usual, $N(\epsilon_F)$ is the electronic 
density of states at the Fermi level.

Following Allen and Dynes \cite{Allen-Dynes75-Tc-S-C} we introduce a two-square-well model

\begin{equation}
    \lambda_\ell(i\omega_m-i\omega_n)= \begin{cases}
               \lambda_\ell        & \mbox{for} \quad |\omega_m|,|\omega_n|\ll\omega_c,\\
               0               & otherwise
           \end{cases}
\end{equation}
with the coupling strength given by 
\begin{equation}
\lambda_\ell=2\int_0^\infty \frac{\omega\; \alpha^2{\cal{F}_\ell}(\omega)}{\omega_{opt}^2+\omega^2}d\omega\equiv N(\epsilon_F)V_{ee}(\ell),
\label{Eq.Lambda-Two-Square-Well}
\end{equation}
where $\omega_{opt}$ is an optimal frequency at which $\lambda_\ell$ is maximal. 
Some comments are in order. The traditional choice for the electron-phonon coupling in the 
two-square-well approximation is $\lambda(i\omega_m-i\omega_n)=\lambda(0)=\lambda$,
where the Matsubara sums are performed, while maintaining $\omega_m=\omega_n$ at
all times. This choice of an instantaneous interaction works well for weakly-coupled 
superconductors, which are characterized by a vibrational spectrum heavily weighted at high 
frequencies in such a way that the characteristic phonon frequency, $\langle\omega\rangle_{\alpha^2{\cal F}}$, 
corresponding to an average over the material's vibrational spectral function, is always much 
higher than any possible difference $|\omega_m-\omega_n|=2\pi |m-n| k_B T/\hslash \ll\omega_c$. 
As spectral weight is transferred towards lower frequencies, however, retardation demands 
that we allow for $m\neq n$ in the Matsubara sums. One could choose $m-n=1$, which 
would result in $\omega_{opt}=2\pi k_B T_c /\hslash$, at the transition temperature, or one could use 
Carbotte's argument for an estimate of the optimal frequency \cite{Carbotte90-SCs-Boson-Exchanged}:
{\it Consider a harmonic oscillator of frequency $\omega$, wherein the polarization is maximal around 
$\omega\approx\omega_{opt}$. The oscillator is then set to oscillate by a passing electron with Fermi 
velocity $v_F$. If the lattice oscillates too slowly, $\omega\ll\omega_{opt}$, there will be no polarization 
effects, within a region of the size of the coherence length, $\xi_0$, on a second passing electron with 
the same velocity $v_F$; if it oscillates too rapidly, $\omega\gg\omega_{opt}$, the polarization will average 
out to zero before the second electron has left the coherence perimeter. Either way, the retarded interaction 
must vanish at both $\omega\rightarrow 0$ and $\omega\rightarrow\infty$ and be maximal at $\omega_{opt}$, 
which in terms of $v_F$ and $\xi_0$ can then be written as $\omega_{opt}=\pi v_F/2\xi_0$.} Alternatively, 
$\omega_{opt}$ has been estimated, after inclusion of various Matsubara frequencies, all satisfying 
$|\omega_m|,|\omega_n|\ll\omega_c$, to be roughly $\omega_{opt}\approx 10 k_B T^{max}_c /\hslash$ 
(slightly higher than the lowest bound of $2\pi k_B T_c /\hslash$ for $m-n=1$ discussed before) where 
$T^{max}_c$ is the maximum possible value for $T_{c}$ in the defectal-free system. All in all, the 
presence of an optimal frequency, $\omega_{opt}\neq 0$, implies that our retarded, effective $V_{ee}(\ell)$ is most effective around $\omega_{opt}$, while drops to zero at both 
$\omega\rightarrow\infty$ and $\omega\rightarrow0$ limits: as also shown in Fig.~\ref{Fig5-FL-Mechanisms-Eff-Vee}(h). 

Using the above notation, $Z_\ell=1+\lambda_\ell$ and the zero gap limit, 
$\Delta_\ell\rightarrow 0$, of Eliashberg's equations reduces to
\begin{equation}
1=\frac{\lambda_\ell-\mu^*}{1+\lambda_\ell}\pi T_c\sum_{|\omega_m|<\omega_c}
\frac{1}{|\omega_m|}\simeq \frac{\lambda_\ell-\mu^*}{1+\lambda_\ell}
\ln{\left[\frac{2e^\gamma \omega_c}{\pi T_c}\right]},
\end{equation}
where $\gamma\simeq 0.577$ is Euler's constant. 
Now we can exponentiate both sides to arrive at
the usual MacMillan \cite{Mcmillan68-Tc} (or simplified Allen-Dynes \cite{Allen-Dynes75-Tc-S-C}) result for $T_{c}$
\begin{equation}
T_c(\ell) = \theta e^{-(1+\lambda_\ell)/(\lambda_\ell-\mu^*)}\quad\mbox{with}\quad
\theta=\frac{1.13 \hbar\omega_c}{k_B }.
\label{Eq.Tc-McMillan}
\end{equation}

\begin{center}
\begin{figure*}[hbtp]
\includegraphics[scale=0.48]{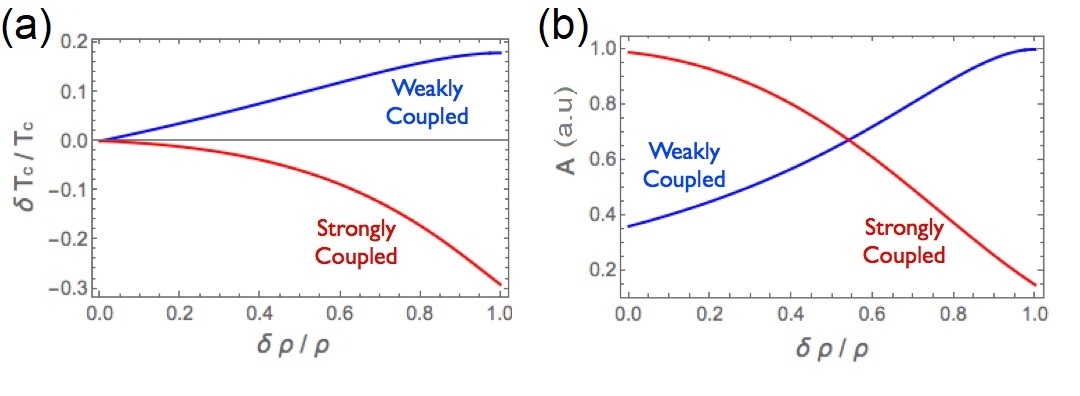}
\caption{Correlated variations of $\delta T_c(\delta\rho_\circ/\rho_\circ)/T_c$ (left, (a)) and 
$A(\delta\rho_\circ/\rho_\circ)$ (right, (b)), 
for the two classes of conventional superconductors considered in this work: 
weakly-coupled superconductors (red, decreasing) and strongly-coupled superconductors (blue, increasing).
}
\label{Fig4-Correlated-Variation-Tc-A}
\end{figure*}
\end{center}

The calculation of $T_c(\ell)$ depends on the strength of the electron-phonon coupling  
$\lambda_\ell$ which, in turn, is determined by the amplitude, $V_{ee}(\ell)$, via
eq. (\ref{Eq.Lambda-Two-Square-Well}). As we can see, $\lambda_\ell$ is highly sensitive to 
{\it softening}, namely, to shifts in spectral weight relative to an optimal frequency $\omega_{opt}$. 
According to the scaling theorem of Coombes and Carbotte \cite{Carbotte90-SCs-Boson-Exchanged}, when the total integrated 
area under the spectral function, $\alpha^{2}\mathcal{F}_\ell \left(\omega\right)$  in Eq.(\ref{Eq-Eliashberg-Spect-Function}), is equal to a constant $\mathcal{A}$, then the best 
shape that maximizes $T_c$ is a $\delta$-function which, here, is introduced as  
an Einstein spectrum
\begin{eqnarray}
\alpha^{2}\mathcal{F}_\ell\left(\omega\right)&=&\mathcal{A}\delta\left(\omega-\omega_{E}(\ell)\right),\nonumber\\
\omega_E(\ell)&\approx&\left(1-\frac{1}{k_F\ell}\right)\omega_E(\infty)+\frac{1}{k_F\ell}\omega_R,
\quad\mbox{for } k_F\ell\gg 1,
\label{Eq-Spectral-Function-Einstein}
\end{eqnarray}
wherein the material dependent $\omega_{E}(\infty)$ is an average phonon frequency calculated 
self-consistently in terms of the defectal-free $\alpha^{2}\mathcal{F}_\infty\left(\omega\right)$; 
$\omega_E(\ell)$ is a monotonically decreasing function of $\ell$, valid for $k_F\ell\gg 1$ as 
$\ell$ is decreased towards $a_\circ$: such a softening is manifested in all materials undergoing amorphization \cite{Bergmann76-SC-AmorphousMetals-Review}. 
The evolution of the normalized $\lambda_\ell$ undergoing such a softening is
\begin{eqnarray}
\frac{\delta\lambda_\ell}{\lambda_\infty}
&\equiv& \frac{\lambda_\ell - \lambda_\infty}{\lambda_\infty}\nonumber\\
&=&\frac{1}{k_F\ell}\left[\frac{\omega_E^2(\infty)}{\omega_{opt}^2+\omega_E^2(\infty)}
\left(2-\frac{1}{2k_F\ell}\right)-1\right]+{\cal O}\left(\frac{1}{\ell^3}\right).
\label{Eq-Delta-Lambda}
\end{eqnarray}
wherein both $\omega_E(\ell)$ and $\omega_{opt}$ are much higher than the resonance frequency $\omega_R$, see 
Fig.~\ref{Fig5-FL-Mechanisms-Eff-Vee}(g)-(h). The associated evolution of
$T_c(\ell)$ is shown in Fig.~\ref{Fig4-Correlated-Variation-Tc-A}(a) for
the two classes of conventional superconductors discussed below.
%

\subsubsection{Weakly coupled superconductors}

A defectal-free member of this class is characterized by $\omega_E(\infty) \gg \omega_{opt},\omega_R$: then $\lambda_\infty\ll 1$ and $T_c$ is low. Incorporation of defectals leads to $\delta\rho_\circ\equiv (\partial\rho_\circ/\partial\ell)\delta\ell>0$ (since $\delta\ell <0$) and, based on Eq.(\ref{Eq-Spectral-Function-Einstein}), a shift in $\omega_E(\ell)$ towards 
$\omega_{opt}$. From the structure of $V_{ee}\left(\omega _E/\omega_{opt}\right)$, shown
in Fig.~\ref{Fig5-FL-Mechanisms-Eff-Vee}(g), one concludes that weakly-coupled superconductors 
exhibit an {\it enhancement} of both $\lambda_\ell$ and $T_c(\ell)$ upon 
defectal incorporation. Furthermore, recalling that $\omega_E(\ell)\gg\omega_{opt},\omega_R$, for 
$1\ll k_F\ell<\infty$, and $\ell/\ell_\circ=\rho_\circ^\circ/\delta\rho_\circ$, one arrives at [see Eq.(\ref{Eq-Delta-Lambda}, \ref{Eq.Tc-McMillan})]
\begin{equation}
\frac{\delta T_c(\delta\rho_\circ/\rho_\circ^\circ)}{T_c^\circ}
\approx\frac{\lambda_\infty}{(\lambda_\infty-\mu^*)^2}
\frac{1}{k_F \ell_\circ}\left(\frac{\delta\rho_\circ}{\rho_\circ^\circ}\right),
\label{Eq-Tc-vs-rho-Low-Coupling}
\end{equation}
for weak but positive
$\frac{\delta\rho_\circ}{\rho_\circ^\circ}$,
a {\it linear evolution universally valid within the weak defectal-concentration range}:  
consistent with the empirical analysis provided earlier 
for superconductors belonging to the upper quadrant of Fig.~\ref{Fig2-Exp-dTc-dRo-BCS}, and with a slope 
determined solely by material properties such as $k_F \ell_\circ$, 
$\lambda_\infty$, and $\mu^*$, see Fig. \ref{Fig4-Correlated-Variation-Tc-A}(a)(blue).
  
\subsubsection{Strongly-coupled  superconductors} 

\begin{center}
\begin{figure*}[hbtp]
\includegraphics[scale=0.35]{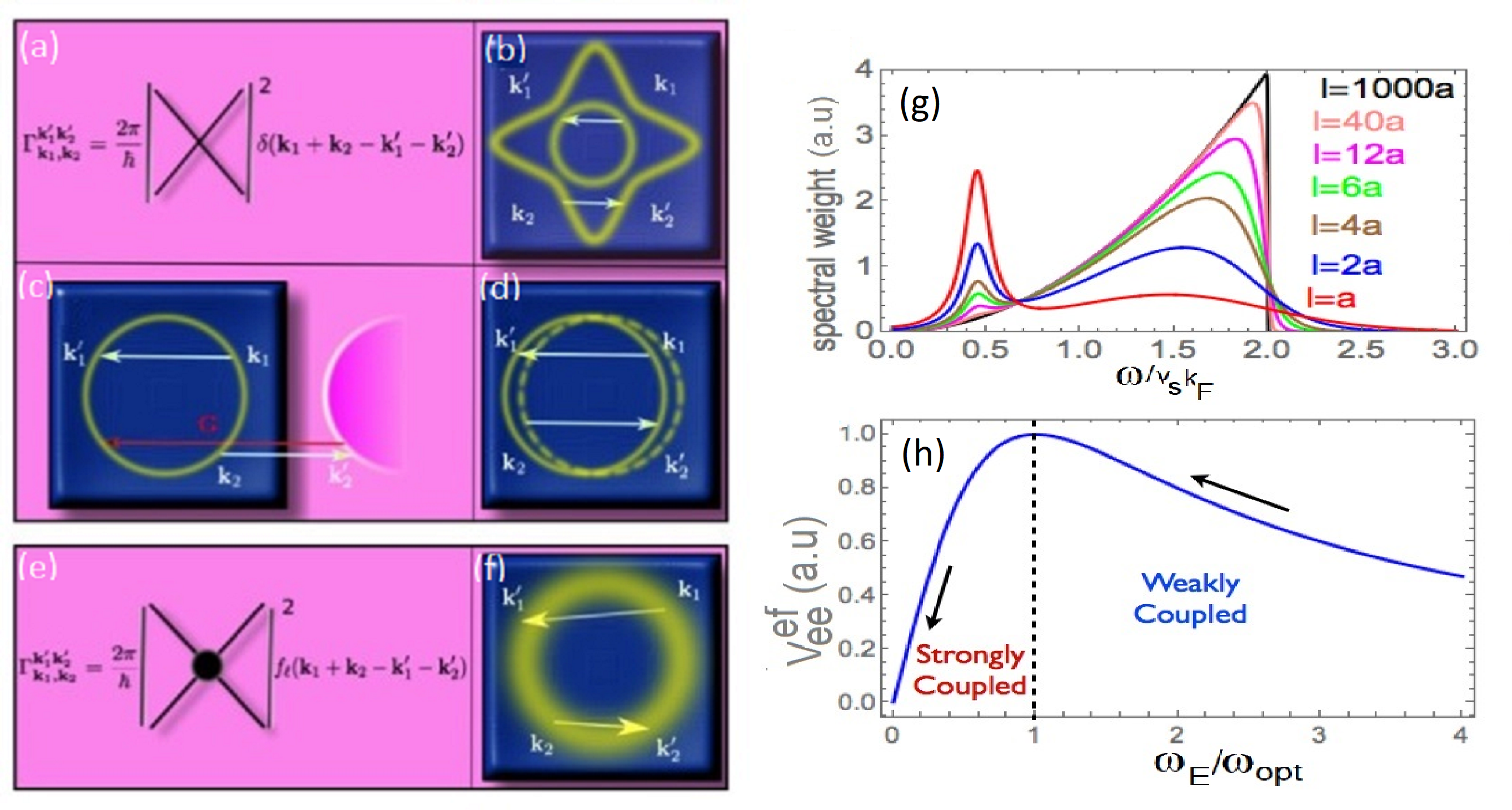}
\caption{
(a) $\Gamma_{{\bf k}_1,{\bf k}_2}^{{\bf k}_1^\prime,{\bf k}_2^\prime}$ of a defectal-free system 
with kinematically constrained relaxation channels: (b) multi-band or Baber scattering; (c) multi-zone, 
or Umklapp, scattering; (d) multi-sheet, or topological, scattering.
(e) $\Gamma_{{\bf k}_1,{\bf k}_2}^{{\bf k}_1^\prime,{\bf k}_2^\prime}$, of a defectal-bearing system
with kinematically unconstrained relaxation channel: (f) the novel {\it halo-umklapp scattering mechanism};
(g) Eliashberg's spectral function, $\alpha^2{\cal F}$ as a funcion of $\omega/v_s k_F$ for 
different values of $\ell$ showing the transfer of spectral weight from high- to low- frequencies;
(h) effective electron-electron, $V_{ee}$, interaction as a function of $\omega_E/\omega_{opt}$ 
showing phonon softening and the increase of the interaction for
weakly-coupled superconductors, and its decrease for strong coupling.
}
\label{Fig5-FL-Mechanisms-Eff-Vee}
\end{figure*}
\end{center}

Here a defectal-free member is characterized 
by $\omega_E(\infty) \lessapprox \omega_{opt}$ and $\omega_E(\infty) \gg \omega_R$; accordingly $\lambda_\infty\sim 1$ and relatively high $T_c$. 
Incorporation of defectals 
in such systems increases $\delta\rho_\circ>0$ and a 
shift in $\omega_E(\ell)$, away from $\omega_{opt}$ but towards $\omega_R$. 
From the structure of $V_{ee}\left(\omega _E/\omega_{opt}\right)$, 
shown in Fig.~\ref{Fig5-FL-Mechanisms-Eff-Vee}(g), one concludes that strongly-coupled 
superconductors exhibit a {\it reduction} of both $\lambda_\ell$ and $T_c(\ell)$ 
upon defectal incorporation. It is recalled that $\omega_E(\ell)\lessapprox\omega_{opt}$ 
and $\lambda_\infty \gg\mu^*$ for $1\ll k_F\ell<\infty$. Then based on 
Eqs.(\ref{Eq-Spectral-Function-Einstein},\ref{Eq-Delta-Lambda}), one obtains
\begin{equation}
\frac{\delta T_c(\delta\rho_\circ/\rho_\circ^\circ)}{T_c^\circ}\approx
-t_1\left(\frac{\delta\rho_\circ}{\rho_\circ^\circ}\right)-t_2\left(\frac{\delta\rho_\circ}{\rho_\circ^\circ}\right)^2,
\label{Eq-Tc-vs-rho-Strong-Coupling}
\end{equation}
with $t_1=(\lambda_\infty^{-1}/k_F \ell_\circ)(1-2\omega_E^2(\infty)/(\omega_{opt}^2+\omega_E^2(\infty)))$, positive
for $\omega_E(\infty)\lessapprox\omega_{opt}$, and a reasonably large value for
$t_2=(\lambda_\infty^{-1}/8 k_F \ell_\circ)\omega_E^2(\infty)/(\omega_{opt}^2+\omega_E^2(\infty))>0$. This shows a  
deviation from linearity which is consistent with the empirical analysis of the lower quadrant of 
Fig.~\ref{Fig2-Exp-dTc-dRo-BCS}, see also Fig. \ref{Fig4-Correlated-Variation-Tc-A}(a)(red).

\subsection{$A(\ell)$ from Boltzmann's transport theory}

The calculation of the FL coefficient $A$ depends on the availability of momentum relaxation channels associated 
with the two-particle process within a defectal-bearing lattice; such an availability is determined 
by the amount of {\it distortion} which directly affects the phase, $\phi_{{\bf q}}$, via
\begin{equation}
f_\ell({\bf k}_1^\prime,{\bf k}_1,{\bf k}_2^\prime,{\bf k}_2)=
\sum_{\bf q}\overline{S}^\ell_{{\bf q}}({\bf k}_1^\prime-{\bf k}_1)\overline{S}^\ell_{-{\bf q}}({\bf k}_2^\prime-{\bf k}_2),
\label{Kinematic-Constraints}
\end{equation}
with the quasi-momentum transfer being controlled by a convolution between 
two $\overline{S}^\ell_{{\bf q}}({\bf k}_i^\prime-{\bf k}_i)$ factors. 
The Fermi liquid coefficient, $A(\ell)$, of a distorted lattice will be calculated from the electron-electron scattering including both the direct Coulomb $V_C$ and the 
retarded, effective (phonon-mediated) electron-electron interaction $V_{ee}(\ell)$. We will look for a variational solution, $\Phi_{{\bf k}}$, to the linearized Boltzmann's 
transport equations in terms of which the resistivity can be written as \cite{Ziman79-Electrons-Phonons}
\begin{equation}
\rho_{ee}(T,\ell)=\frac{\langle\Phi_{{\bf k}},{\cal P}_\ell\Phi_{{\bf k}}\rangle}
{\left| \langle\Phi_{{\bf k}},X\rangle\right|^2}=A(\ell)T^2,
\end{equation}
where ${\cal P}_\ell$ is the {\it scattering operator} that transforms the variational solution 
$\Phi_{{\bf k}}$ into another momentum state, ${\bf k}^\prime$. Through integration, the normalization factor reads
\begin{equation}
\left|\langle\Phi_{{\bf k}},X\rangle\right|=\left| 2e\sum_{{\bf k}}v_{{\bf k}}\Phi_{{\bf k}}\left(-\frac{\partial f\left(\epsilon_{{\bf k}}\right)}{\partial\epsilon_{{\bf k}}}\right)\right|=\frac{e k_F^3}{3\pi^2\hbar m^*}=\frac{n e}{m^*},
\end{equation}
with $\Phi_{{\bf k}}=\vec{u}\cdot\vec{{\bf v}}_{{\bf k}}$ measuring the deviation of the electronic distribution from equilibrium, 
$\vec{u}$ being the direction of the applied electric field, and $\vec{{\bf v}}_{{\bf k}}$ 
the velocity of the quasiparticle associated with momentum ${\bf k}$. In
terms of these quantities, the numerator reads \cite{Ziman79-Electrons-Phonons}
\begin{widetext}
\begin{eqnarray}
\langle\Phi_{{\bf k}},{\cal P}_\ell\Phi_{{\bf k}}\rangle  = 
\frac{1}{2k_{B}T} &\sum_{{\bf k}_1, {\bf k}_2, {\bf k}_1^\prime, {\bf k}_2^\prime}&
f_{\epsilon_{{\bf k}_{1}}}f_{\epsilon_{{\bf k}_{2}}}(1-f_{\epsilon_{{\bf k}_{1}^{\prime}}})(1-f_{\epsilon_{{\bf k}_{2}^{\prime}}})
\delta(\epsilon_{{\bf k}_1}+\epsilon_{{\bf k}_2}-\epsilon_{{\bf k}_1^\prime}-\epsilon_{{\bf k}_2^\prime})\nonumber\\
&\times&
(\Phi_{{\bf k}_{1}}+\Phi_{{\bf k}_{2}}-\Phi_{{\bf k}_{1}^{\prime}}-\Phi_{{\bf k}_{2}^{\prime}})^{2}
\Gamma_{{\bf k}_{1}+{\bf k}_{2}\rightarrow {\bf k}_{1}^{\prime}+{\bf k}_{2}^{\prime}}(\ell),
\label{Eq-Numeral-Boltzmann}
\end{eqnarray}
\end{widetext}
wherein $\Gamma_{{\bf k}_{1}+{\bf k}_{2}\rightarrow {\bf k}_{1}^{\prime}+{\bf k}_{2}^{\prime}}(\ell)$ 
is the transition amplitude for the total electron-electron interaction, 
$V_{tot}(\ell)$, while $f_{\epsilon_{{\bf k}}}$ is the equilibrium electron distribution
\begin{equation}
f_{\epsilon_{{\bf k}}}=\frac{1}{e^{\beta(\epsilon_{{\bf k}}-\mu)}-1},
\end{equation}
with $\beta=1/k_{B}T$, $\epsilon_F$ the Fermi energy and $\mu=\epsilon_{F}$. 

At this point it is important to emphasize that when approaching the superconducting instability 
from the Fermi-liquid state, $T\rightarrow T_c^+(\ell)$, the total electron-electron interaction, 
$V^0_{tot}(\ell)=V_{C}-V_{ee}(\ell)$ becomes renormalized according to
\begin{equation}
V^0_{tot}(\ell)\rightarrow V_{tot}(\ell)=\frac{V_C-V_{ee}(\ell)}{Z_\ell}=\frac{V_C-V_{ee}(\ell)}{1+\lambda_\ell},
\label{Eq-Renormalized-Vtot(l)}
\end{equation}
where the renormalization constant, $Z_\ell$, is the same as obtained when
approaching from the superconducting ground state, $T\rightarrow T_c^-(\ell)$. This is an 
asymptotically exact result obtained by using a renormalization group procedure that treats 
the direct and effective parts of the total interaction on equal footing \cite{Tsai05-Renormalization-Strong-coupled-SUCs}.

At low temperatures, we can project all electron states at the Fermi surface, 
$|{\bf k}_i|=k_F$, and transform $\sum_{{\bf k}_i}$ into integrals over 
$\epsilon_{{\bf k}_i}$, with a constant electronic density of states 
at the Fermi level, as well as integrals over solid angles. The constraint of 
energy conservation can also be rewritten in terms of the transferred energy, 
$\hbar\omega$, to the phonons
\begin{equation} 
\delta(\epsilon_{{\bf k}_1}+\epsilon_{{\bf k}_2}-\epsilon_{{\bf k}_1^\prime}-\epsilon_{{\bf k}_2^\prime})=
\int_{-\infty}^{\infty} d\omega \;  \delta(\epsilon_{{\bf k}_1^\prime}-\epsilon_{{\bf k}_1}-\hbar\omega) \delta(\epsilon_{{\bf k}_2^\prime}-\epsilon_{{\bf k}_2}+\hbar\omega),
\end{equation}
which allows us to eliminate both $\epsilon_{{\bf k}_1^{\prime}}$ and $\epsilon_{{\bf k}_2^{\prime}}$, 
after which we are left with
\begin{widetext}
\begin{equation}
\int d\epsilon_{{\bf k}_{1}}\int d\epsilon_{{\bf k}_{2}}\:
f_{\epsilon_{{\bf k}_{1}}}f_{\epsilon_{{\bf k}_{2}}}
(1-f_{\epsilon_{{\bf k}_{1}}+\hbar\omega})(1-f_{\epsilon_{{\bf k}_{2}}-\hbar\omega})
=\frac{\hbar^{2}\omega^{2}}{(e^{\beta\hbar\omega}-1)(1-e^{-\beta\hbar\omega})}.
\end{equation}
\end{widetext}
Finally, we can integrate over the transferred energy $\hbar\omega$ to obtain
\begin{equation}
\frac{1}{2 k_B T}\int_{-\infty}^{\infty}d(\hbar\omega)
\frac{\hbar^{2}\omega^{2}}{(e^{\beta\hbar\omega}-1)(1-e^{-\beta\hbar\omega})}=
\frac{\pi^{2}}{3}(k_{B}T)^{2},
\label{Eq.R-prop-T2}
\end{equation}
which ensures that the contribution to the resistivity from this novel electron-electron 
interaction has the typical Fermi liquid {\it quadratic-in-T}, $\rho_{ee}(T,\ell)=A(\ell) T^2$ with
\begin{widetext}
\begin{equation}
A(\ell)=\left(\frac{m^*}{ne}\right)^2\frac{\pi^2 k_B^2}{3(\hbar v_F)^4}
\int\int\int\int\frac{d\Omega_{{\bf k}_1}d\Omega_{{\bf k}_1^\prime}
d\Omega_{{\bf k}_2}d\Omega_{{\bf k}_2}^\prime}{(2\pi)^{12}}
(\Phi_{{\bf k}_{1}}+\Phi_{{\bf k}_{2}}-\Phi_{{\bf k}_{1}^{\prime}}-\Phi_{{\bf k}_{2}^{\prime}})^{2}\times
\Gamma_{{\bf k}_{1}+{\bf k}_{2}\rightarrow {\bf k}_{1}^{\prime}+{\bf k}_{2}^{\prime}}(\ell),
\label{Eq.A-N.Gamma-Vee^2}
\end{equation}
\end{widetext}
and $\Gamma_{{\bf k}_{1}+{\bf k}_{2}\rightarrow {\bf k}_{1}^{\prime}+{\bf k}_{2}^{\prime}}(\ell)$ 
calculated by the use of Fermi's golden rule
\begin{equation}
\Gamma_{{\bf k}_{1}+{\bf k}_{2}\rightarrow {\bf k}_{1}^{\prime}+{\bf k}_{2}^{\prime}}(\ell)=
\left(\frac{2\pi}{\hbar}\right)\left|V_{tot}(\ell)\right|^2 f_\ell({\bf k}_1+{\bf k}_2-{\bf k}_1^\prime-{\bf k}_2^\prime),
\label{Eq.Gamma-Vee^2}
\end{equation}
where $f_\ell({\bf k}_1+{\bf k}_2-{\bf k}_1^\prime-{\bf k}_2^\prime)$ regulates all 
kinematic constraints through eq. (\ref{Kinematic-Constraints}).
After projecting all ${\bf k}_i$ states onto the roughened Fermi surface, 
we thus obtain 
\begin{equation}
A(\ell)=F^2_\ell\left|V_{tot}(\ell)\right|^2,
\label{FL-Coeff-A}
\end{equation}
where the so called {\it efficiency of momentum relaxation} 
\begin{widetext}
\begin{equation}
F^2_\ell=F^2_\circ\int\int\int\int\frac{d\Omega_{{\bf k}_1}\, d\Omega_{{\bf k}_1^\prime}\, d\Omega_{{\bf k}_2}\, d\Omega_{{\bf k}_2^\prime}}{(2\pi)^{12}}
(\Phi_{{\bf k}_{1}}+\Phi_{{\bf k}_{2}}-\Phi_{{\bf k}_{1}^{\prime}}-\Phi_{{\bf k}_{2}^{\prime}})^{2}\times
f_\ell({\bf k}_1+{\bf k}_2-{\bf k}_1^\prime-{\bf k}_2^\prime),
\label{Eq-Fell}
\end{equation}
\end{widetext}
with
\begin{equation}
F^2_\circ=(2\pi/\hbar)(m^*/ne)^2(\pi^2 k_B^2/3 \hbar^4 v_F^4)(1/N^2(\epsilon_F)),
\label{Eq-F_o}
\end{equation}
where $\Phi_{{\bf k}}=\vec{u}\cdot {\bf v}_{\bf k}$,
$\vec{u}$ is the unit vector along the direction of the applied electric field, 
${\bf v}_{\bf k}=\hbar {\bf k}/m^*$ is the quasiparticle velocity of carriers having effective mass $m^*$.

\subsubsection{Relaxed kinematics and the robustness of the FL}

The precise evaluation of $F^2_\ell$ requires a microscopic calculation that includes all possible 
relaxation, momentum-transferring channels such that
\begin{equation}
[\vec{u}\cdot({\bf k}_{1}+{\bf k}_{2}-{\bf k}_{1}^{\prime}-{\bf k}_{2}^{\prime})]^2\times f_\ell({\bf k}_1+{\bf k}_2-{\bf k}_1^\prime-{\bf k}_2^\prime)
\neq 0.
\end{equation}
As evident, the fate of the Fermi-liquid coefficient, $A(\ell)$, for different Fermi surface topologies, 
will be determined essentially by the available phase space for scattering, because this
enters into the integration over solid angles, $\int...d\Omega_{{\bf k}_i}$, with the integrand containing 
the product between the large angle scattering factor, equivalent to $(1-\cos\theta)$, and the scattering transition amplitude, 
$(\Phi_{{\bf k}_{1}}+\Phi_{{\bf k}_{2}}-\Phi_{{\bf k}_{1}^{\prime}}-\Phi_{{\bf k}_{2}^{\prime}})^{2}
\times f_\ell({{\bf k}_{1}+{\bf k}_{2}\rightarrow {\bf k}_{1}^{\prime}+{\bf k}_{2}^{\prime}})$. 

In a defectal-free system (e.g. our example of Al thin films), we have 
$f_\infty({{\bf k}_{1}+{\bf k}_{2}\rightarrow {\bf k}_{1}^{\prime}+{\bf k}_{2}^{\prime}})=
\delta_{{{\bf k}_{1}+{\bf k}_{2},{\bf k}_{1}^{\prime}+{\bf k}_{2}^{\prime}}}$, and kinematics 
severely restricts the availability of phase space for net momentum transfer to the lattice to
the following scattering channels shown in Fig. \ref{Fig5-FL-Mechanisms-Eff-Vee}(b)-(d): 
(i) the Baber mechanism, for a multi-band 
Fermi surface \cite{Baber37-AT2-contribution}; (ii) the umklapp mechanism, for Fermi surfaces that are at least quarter-filled \cite{Yamada86-FL-using-Anderson-Hamiltonion,Maebashi98-Umklamp-Resis-FL}; 
and (iii) the normal mechanism, for multiply connected Fermi surfaces with an infinite number 
of self-intersecting points \cite{Pal-Maslov-e-e-FL-theory}. As a result, any $A(\infty)T^2$ 
contribution allowed by these channels is typically very small, 
$A(\infty)\sim 10^{-7}\mu\Omega$cmK$^{-2}$ (low scattering efficiency), or even 
identically vanishing, $A(\infty)\equiv 0$, for topologically trivial, single band, 
small Fermi surface systems. 
The extreme specificity of the above relaxation mechanisms is in stark disagreement with the ubiquitous 
experimental observation of a robust Fermi liquid behavior in the transport properties of defectal-bearing
superconductors (see, e.g., Fig.~\ref{Fig3-Exp-Tc-A-Ro-Exp-lnTc-sqrA}).

The stabilization of the FL behaviour
clearly requires a remarkable increase in phase space for scattering as the one promoted by 
defectals. In this case, the kinematic constraints are relaxed due to the breakdown of translational invariance,
$f_\ell({{\bf k}_{1}+{\bf k}_{2}\rightarrow {\bf k}_{1}^{\prime}+{\bf k}_{2}^{\prime}})\neq
\delta_{{{\bf k}_{1}+{\bf k}_{2},{\bf k}_{1}^{\prime}+{\bf k}_{2}^{\prime}}}$,
and a robust FL behaviour results, $A(\ell<\infty)\neq 0$, 
over a rather wide temperature range. We call the mechanism behind 
such large values for $A(\ell<\infty)\neq 0$ as {\it halo-umklapp scattering}, 
whereby the enlargement of the available phase space for quasi-momentum 
relaxation stems from the monotonically increasing broadening of the Bragg 
reflections at higher-order reciprocal lattice points, which eventually merge 
together to produce a {\it halo-shaped} diffraction pattern as the one shown
in Fig. \ref{Fig1-Point-Defectals-Model}(k).

\subsubsection{$A(\ell)$ for weakly- and strongly- coupled superconductors}

The two-electron scattering mechanism on a {\it distorted Fermi surface}
shown in Figs.~\ref{Fig1-Point-Defectals-Model}(e,g,i) and Figs. \ref{Fig5-FL-Mechanisms-Eff-Vee}(e,f)
is denoted as {\it halo-umklapp mechanism}. The associated value for $A(\ell)$
is given by Eq.(\ref{FL-Coeff-A}) and plotted in Fig.~\ref{Fig4-Correlated-Variation-Tc-A}(b)
as a function of $\delta\rho_\circ/\rho_\circ$: the red (blue) lines represent its evolution for 
strongly- (weakly)-coupled superconductors.
For small $\delta\rho_\circ$ and $1\ll k_F\ell <\infty$ we obtain, for $A(\delta\rho_\circ)$
\begin{equation}
A(\ell)=A(\infty) + a_1 \delta\lambda_\ell + a_2 (\delta\lambda_\ell)^2 =
A_\circ + a_3 \delta\rho_\circ + a_4 (\delta\rho_\circ)^2 ,
\label{Eq-Avsrho}
\end{equation}
where $A(\infty)$ and $A_\circ$ refer to the negligibly small kinematically-constrained 
contributions from the host 
matrix, while $a_1=2(|\lambda_\infty-\mu^*|(1+\mu^*))/(1+\lambda_\infty)^3$ as well as
$a_2=(1+\mu^*-|\lambda_\infty-\mu^*|)(1+\mu^*)/(1+\lambda_\infty)^4$,
include all kinematically unconstrained relaxation processes following defectal incorporation. 
For empirical results, see Figs.~\ref{Fig3-Exp-Tc-A-Ro-Exp-lnTc-sqrA}(a.8,b.1).

\section{Reconciling $T_c(\ell)$ with $A(\ell)$}

We are now ready to unveil the mechanism that bridges $T_c$ and $A$ in terms of a
universal kinematic correlation with the experimentally observed 
BCS-like form $T_c(\ell)/\theta=e^{-F_\ell/\sqrt{A(\ell)}}$. For that purpose, let 
us recall two features: (i) due to the relaxation of the kinematic constraints, $A(\ell)$ is 
proportional to the square of the electronic density of states at the Fermi level 
\begin{equation}
A(\ell)\propto\frac{\sum_{{\bf k}_i,i=1\dots 4}}{|\sum_{{\bf k}}|^2}\rightarrow N^2(\epsilon_F);
\end{equation}
(ii) on approaching the superconducting instability from the Fermi-liquid state,
the total $V_{tot}(\ell)$ interaction is also renormalized, as in Eq.(\ref{Eq-Renormalized-Vtot(l)}),
$\lambda_\ell=N(\epsilon_F)V_{ee}(\ell)$ and also, $\mu=N(\epsilon_F)V_C$, with
the pseudopotential itself renormalized as $\mu^*=\mu/(1+\mu\ln{(\epsilon_F/\omega_c)})$,
resulting in
\begin{equation}
A_{\ell}\propto N(\epsilon_F)^2|V_{tot}(\ell)|^2 = F_\ell^2 \left(\frac{\lambda_\ell - \mu^*}{1+\lambda_\ell}\right)^2.
\end{equation}
Finally, once we recall Eliashberg's result for the critical temperature it is straightforward to conclude that
\begin{equation}
T_c (\ell) = 1.13 \frac{\hbar\omega_c}{k_B} e^{-(1+\lambda_\ell)/(\lambda_\ell - \mu^*)}=\theta e^{-F_{\ell}/\sqrt{A_{\ell}}},
\label{Tc-of-A}
\end{equation}
wherein $\theta$ is as given in Eq.(\ref{Eq.Tc-McMillan})
and $F^2_\ell$, as in Eq.(\ref{Eq-Fell}).
 
Within the spirit of the renormalization group, 
defectal incorporation  in conventional superconductors can be seen as a relevant perturbation that promotes the
running of $\lambda_\ell$ towards either the weak- or strong-coupling limits, depending on
the relative values of $\omega_E(\infty)$ and $\omega_{opt}$. For weakly-coupled
superconductors, where $\omega_E(\infty)\gg\omega_{opt}$ and $\lambda_\infty\ll 1$, 
the running is towards stronger couplings, while for the strongly-coupled superconductors, 
where $\omega_E(\infty)\lessapprox\omega_{opt}$ and $\lambda_\infty\approx 1$, the 
running is towards weaker couplings. If we recall the relation between $T_c(\ell)$ and $A(\ell)$
given in eq. (\ref{Tc-of-A}) we conclude that the incorporation of defectals promotes the 
{\it correlated flow of $T_c(\ell)$ and $A(\ell)$, without ever leaving the curve defined 
by Eq.(\ref{Tc-of-A})}.

\section{Discussions and Outlook\label{Sec.Conclusions}}

The Kadowaki-Woods ratio is defined as 
$A/\gamma^2$ which is expected to be a universal 
constant in Fermi liquids since $A\propto {m^*}^2$ and $\gamma\propto m^*$. 
For defectal-related electron-phonon or spin-fluctuation Fermi-liquids, we predict that the Kadowaki-Woods ratio should
be larger by a geometric factor $F^2_\ell/F^2_\infty$
\begin{equation}
\frac{A(\ell)}{\gamma^2}=\frac{81}{4\pi\hbar k_B^2 e^2}\left(\frac{F^2_\ell}{F^2_\infty}\right)
\frac{1}{d^2 n N^2(\epsilon_F)\langle v_{0x}^2\rangle}.
\label{KDW-ratio-Defectal}
\end{equation}
where $\langle v_{0x}^2\rangle$ is a Fermi surface average of the carrier velocity squared
that accounts for anisotropies, $e$ is the electric charge of the direct, Coulomb, electric-electric
interaction, $n$ is the carrier density, and $d \sim 1$ is a dimensionless number. 
As we have discussed earlier, $F^2_{\ell}/F^2_\infty$ is a measure of the efficiency of 
momentum relaxation via umklapp (or any other kind of) scattering and $F^2_\ell > F^2_\infty$ in 
Eq.(\ref{KDW-ratio-Defectal}) as a result of the easing of the kinematic constraints of momentum 
conservation: an intrinsic character of a distorted lattice.  

We also calculated the gap-to-$T_c$ ratio  of a defectal-bearing superconductor, beyond the $\theta/T_c\rightarrow\infty$ approximations
\begin{equation}
\frac{2\Delta(\ell)}{k_B T_c(\ell)}=3.53\left\{1+12.5 \left[\frac{T_c(\ell)}{\theta}\right]^2\ln{\left[\frac{\theta}{2T_c(\ell)}\right]}\right\}.
\end{equation}
This equation indicates that, for weakly-coupled, superconducting, defectal-free simple metals, the gap-to-$T_c$ ratio is the universal ratio
$2\Delta(\infty)/k_B T_c(\infty)= 3.53$. As defectals are incorporated, this ratio increases 
with $T_c(\ell)$, showing that the flow is towards stronger couplings. The opposite 
occurs for the case of defectal-free, strongly-coupled, superconductors, where 
$2\Delta(\infty)/k_B T_c(\infty)= 3.53\{1+12.5 [T_c(\infty)/\theta]^2\ln{[\theta/2T_c(\infty)]}\}>3.53$ is
nonuniversal, but as defectals are incorporated, $T_c(\ell)$ decreases while this ratio decreases, 
towards the universal ratio $3.53$, showing that the flow is towards weaker couplings. 

Finally, as an outlook, it is of extreme interest to extend our analysis 
to other classes of superconducting families not discussed here, such as 
magnesium diboride \ce{MgB2}, which is a strong-coupled,
two-gap superconductor that also manifests a correlated $T_c$ and $A$ \cite{Nunes12-FermiLiquid-SUC}, 
the pervoskite titanate SrTiO$_{3-\delta}$, which manifest a correlated superconductivity and 
$AT^2$ contribution within a range nearing ambient temperature \cite{Lin14-SrTiO3-SC-2bands,Lin15-SrTiO3-AT2-Small-FS}, 
the conventional high $T_c$ sulpher hydride H$_{2}$S superconductor \cite{Drozdov15-sulfurHydride-203K}, 
and the overdoped, high-T$_c$, superconducting cuprates, which manifest a Fermi-liquid regime close to the 
superconducting state \cite{Pines13-SpinFluctuation-SUC}. All of these are well-known for their 
defect-bearing character and in addition each contains the often-anchor-acting hydrogen/oxygen 
as one of the constituent elements.


\section*{Acknowledgements}

We are grateful to Pedro B. Castro and Davi A. D. Chaves for their assistance in the literature 
search and experimental data analysis during the initial stage of this project. The authors also
acknowledge Indranil Paul and Eduardo Miranda for numerous and fruitful discussions. 

\bibliography{FL-SC-Defectal-letter}

\appendix

\section{The el-ph coupling on a distorted lattice \label{El-Ph-Coupling-Defectal}}

The interaction Hamiltonian for a system of band electrons, having a dispersion $\epsilon({\bf k})$,
and coupled to a phonon bath, with dispersion $\omega_{{\bf q},\nu}$, reads  
\begin{equation}
H_{el-ph}=\sum_{\mathbf{k}^{\prime},\mathbf{k},\sigma;\mathbf{q},\nu}
\varphi({\bf k}^\prime-{\bf k}-{\bf q})
g_{\mathbf{k}^{\prime},\mathbf{k},\mathbf{q},\nu}\,
c_{\mathbf{k}^{\prime},\sigma}^{\dagger}c_{\mathbf{k},\sigma}
\left(a_{\mathbf{q},\nu}+a_{-\mathbf{q},\nu}^{\dagger}\right),
\label{Eq-H-ep}
\end{equation}
wherein $c^\dagger_{{\bf k}^\prime,\sigma},c_{{\bf k},\sigma}$ are the creation and annihilation 
operators, respectively, for the fermionic particles with momenta ${\bf k}^\prime,{\bf k}$ and spin $\sigma$;
$a^\dagger_{{\bf q},\nu},a_{{\bf q},\nu}$ are the creation and annihilation operators for phonons with 
momentum ${\bf q}$ at the branches $\nu=L,T_1,T_2$, with polarization unit vector 
$\hat{e}\left(\nu,\mathbf{q}\right)$, and
\begin{equation}
g_{\mathbf{k}^{\prime},\mathbf{k},\mathbf{q},\nu}  =  
\sqrt{\frac{\hbar}{2MN\omega_{\mathbf{q},\nu}V}}
\hat{e}\left(\nu,\mathbf{q}\right)\cdot\left(\mathbf{k}^{\prime}-\mathbf{k}\right)
\label{Eq-g-q-lambda-General}
\end{equation}
is the {\it amplitude} of the electron-phonon matrix element within the deformation potential approximation, 
for $N$ ions of mass $M$ in a volume $V$, while the {\it phase interference factor} is 
\begin{equation}
\varphi({\bf k}^\prime-{\bf k}-{\bf q})=\frac{1}{\sqrt{N}}\sum_{{\bf r}}e^{i({\bf k}^\prime-{\bf k}-{\bf q})\cdot{\bf r}},
\end{equation}
where the sum runs over all $N$ sites of the lattice, ${\bf r}=n_1{\bf a}_{1}+n_2{\bf a}_{2}+n_3{\bf a}_{3}$,
where ${\bf a}_{i=1,2,3}$ are the primitive vectors of the unit cell and $n_{i=1,2,3}$ are integer numbers. 

During scattering, any exchange of momenta between electrons, ${\bf k}, \, {\bf k}^\prime$, 
and phonons, ${\bf q}$, is controlled by the electron-phonon structure factor, 
$S_{{\bf q}}({\bf k}^\prime-{\bf k})$, which, for a pristine, translationally invariant crystal, reduces to
\begin{eqnarray}
S_{{\bf q}}({\bf k}^\prime-{\bf k})&\equiv& |\varphi({\bf k}^\prime-{\bf k}-{\bf q})|^2=
\frac{1}{N}\sum_{{\bf r},{\bf r}^\prime}e^{i({\bf k}^\prime-{\bf k}-{\bf q})\cdot ({\bf r}-{\bf r}^\prime)}\nonumber\\
&=&\sum_{{\bf g}}\delta_{{\bf k}^\prime-{\bf k}-{\bf q},{\bf g}},
\label{Eq-Structure-factor-General}
\end{eqnarray} 
where the sum over the reciprocal lattice points ${\bf g}=h{\bf b}_1+k{\bf b}_2+l{\bf b}_3$ runs over 
all integers $h,k,l$, with ${\bf b}_{i=1,2,3}$ being the primitive vectors of the reciprocal lattice that 
satisfy Laue's condition ${\bf a}_i\cdot{\bf b}_j=2\pi\delta_{ij}$. In this case, quasi-momentum is 
conserved {\it exactly}, as scattering occurs solely for certain allowed values for the 
momentum transfer, ${\bf q}={\bf k}^\prime-{\bf k}-{\bf g}$, as dictated by translational invariance.

The electron-phonon structure factor in Eq.(\ref{Eq-Structure-factor-General}) will certainly
be modified by the distortions associated with the intentional incorporation of defectals. A closer look at Fig.~\ref{Fig1-Point-Defectals-Model}(b) suggests that 
the defectal arrangement can be visualized as a metallic granule dispersed in a perfect 
metallic host. For the purpose of this work, 
we shall introduce the notion of a {\it distorted lattice}, given in 
Fig.~\ref{Fig1-Point-Defectals-Model}(c), where the distortions introduced by 
defectals are taken to be distributed throughout the entire lattice. In this case, 
distortion will be associated to a statistical probability distribution in the coordination 
of the ions in the direct lattice. We follow closely the notation introduced by Hosemann
\cite{Hosemann} for describing diffraction patterns in paracrystals and we introduce a 
Gaussian distribution,
\begin{equation}
P_i({\bf a})=\frac{1}{\sqrt{2\pi}\sigma}e^{-({\bf a}-\overline{\bf a}_i)^2/2\sigma^2},
\end{equation}
whose first three moments are given by
\begin{eqnarray}
\int\, d^3{\bf r}\, P_i({\bf a})&=&1,\nonumber\\
\int\, d^3{\bf r}\, {\bf a}\, P_i({\bf a})&=&\overline{\bf a}_i,\nonumber\\
\frac{1}{\overline{\bf a}_i^2}\int\, d^3{\bf r}\, ({\bf a}-\overline{\bf a}_j,\overline{\bf a}_i)^2\, P_j({\bf a})&=&\sigma_{ij}^2.
\end{eqnarray}
For simplicity, we shall assume the fluctuations within each of the three crystallographic
directions to have the same variance, $\sigma_{ii}=\sigma$, and to be uncorrelated for $i\neq j$. 
In this case, the whole distorted crystal corresponds to a convoluted network of linearly autocorrelated 
lattice positions in which case the distorted structure factor reduces to
\begin{widetext}
\begin{equation}
\overline{S}^\sigma_{\bf q}({\bf k}_1^\prime-{\bf k}_1)=\Pi_{i=1,2,3}
\left\{1+2\,\sum_{n_i=1}^{\infty}\frac{1}{(\sqrt{2\pi n_i}\sigma)^3}
\int d^3 {\bf r}_i \, e^{-({\bf r}_i-n_i\overline{\bf a}_i)^2/2n_i\sigma^2} 
\,e^{i({\bf k}^\prime-{\bf k}-{\bf q})_i {\bf r}_i}\right\}.
\end{equation}
\end{widetext}
The integrals over ${\bf r}_i$ and sums over $n_i$ can be done exactly to produce (for ${\bf p}\equiv{\bf k}_1^\prime-{\bf k}_1-{\bf q}$)
\begin{equation}
\overline{S}^\sigma_{\bf q}({\bf k}_1^\prime-{\bf k}_1)=
\Pi_{i=1,2,3}\left\{\frac{1-|F_i({\bf p})|^2}{(1-|F_i({\bf p})|)^2+4|F_i({\bf p})|\sin^2{[\frac{\overline{\bf a}_i\cdot{\bf p}}{2}]}}\right\},
\label{Structure-Factor-Paracrystal}
\end{equation}
with $F_i({\bf p})$ given, in the Guinier approximation, by
\begin{equation}
F_i({\bf p})=e^{-\sigma^2(\overline{\bf a}_i\cdot{\bf p})^2/2+i\overline{\bf a}_i\cdot{\bf p}}.
\end{equation}
Then the distorted structure factor obtained in Eq. (\ref{Structure-Factor-Paracrystal})
consists of peaks centered at reciprocal lattice positions ${\bf p}\equiv({\bf k}_1^\prime-{\bf k}_1-{\bf q})={\bf g}$, 
with, however, ever decreasing heights given by $S_{max}(m_i)=\overline{a}_i^2/(\pi^2\sigma^2m_i^2)$, where ($m_i=h,k,l$)
are integers, and with ever broader widths $\delta {p}_i=(\pi^2\sigma^2 m_i^2)/\overline{a}_i$. The only true
$\delta$-peak occurs for $h=k=l=0$ and, for that reason, we can make an expansion of the structure factor 
around these maxima, separating the contributions form ${\bf g}=0$ and ${\bf g}\neq 0$ as
\begin{equation}
\overline{S}^\ell_{\bf q}({\bf k}_1^\prime-{\bf k}_1)\approx \delta_{{\bf k}_1^\prime-{\bf k}_1-{\bf q},0}
+\sum_{{\bf g}\neq 0}\frac{S_{max}({\bf g})}{1+\ell^2({\bf k}_1^\prime-{\bf k}_1-{\bf q}-{\bf g})^2}.
\label{Eq-structure-factor-Defectal-S}
\end{equation}
Here, we have $S_{max}({\bf g})=a_\circ^2/\sigma^2\pi^2(h^2+k^2+l^2)$, for $\overline{\bf a}_i=a_\circ$, and 
we have introduced the parameter $\ell \equiv |\delta{\bf g}|^{-1}=a_\circ/\sigma^2\pi^2(h^2+k^2+l^2)$, associated 
with the inverse width of the peaks in the structure factor. Now, quasi-momentum is no longer conserved, in the 
sense that the transferred momentum ${\bf q}={\bf k}_1^\prime-{\bf k}_1-{\bf g}-\delta{\bf g}$ becomes increasingly 
arbitrary, both in magnitude and direction, as $\delta{\bf g}$ becomes larger and larger for higher Brillouin zones, 
until the broadened Bragg reflections merge together producing a {\it halo}, similar to the broadening of the Fraunhoffer 
diffraction pattern observed in crystals with distortions, see Figs.~\ref{Fig1-Point-Defectals-Model}(j-k). 

\section{Eliashberg's spectral function \label{Eliashberg-Spectral-Function-Defectal}}

\subsection{Perfect crystals: $\alpha^{2}\mathcal{F}_{\ell}\left(\omega\right)\sim\omega^2$}

An important quantity in Eliashberg's theory of superconductivity is the spectral function
\begin{widetext}
\begin{equation}
\alpha^{2}\mathcal{F}\left(\omega\right)  = \frac{1}{N(\epsilon_F)} 
\sum_{\mathbf{k}^{\prime},\mathbf{k},\mathbf{q},\nu}
S_{\mathbf{q}}(\mathbf{k}^{\prime}-\mathbf{k})
\left|g_{\mathbf{k}^{\prime},\mathbf{k},\mathbf{q},\nu}\right|^{2}
\delta\left(\epsilon({\bf k})-\epsilon_F \right)\delta\left(\epsilon({\bf k}^\prime)-\epsilon_F \right)
\delta\left(\omega-\omega_{\mathbf{q},\nu}\right).
\label{Eq-Eliashberg-Spectral-Function}
\end{equation}
\end{widetext}
For a defectal-free crystal, $\ell\rightarrow\infty$, the structure factor corresponds to an infinite collection of
$\delta$-peaks, each centered at a ${\bf g}$ in the reciprocal lattice. The interference pattern is clear. For the case of long-wavelength phonons, 
however, we can retain only the ${\bf g}=0$ contribution to $S_{\mathbf{q}}(\mathbf{k}^{\prime}-\mathbf{k})$. In this case, 
$S_{\mathbf{q}}(\mathbf{k}^{\prime}-\mathbf{k})\approx \delta_{\mathbf{k}^{\prime}-\mathbf{k}-\mathbf{q},0}$,
and the above expression for Eliashberg's spectral function reproduces the well known
$\alpha^{2}\mathcal{F}\left(\omega\right)\sim \omega^2$ 
behaviour observed in simple metals with a Debye dispersion, $\omega_{{\bf q},\nu}\propto |{\bf q}|$, 
because ${\bf q}=\mathbf{k}^{\prime}-\mathbf{k}$ exactly and
\begin{eqnarray}
\alpha^{2}\mathcal{F}\left(\omega\right)  &\sim&
\sum_{\left\{\mathbf{k}^{\prime},\mathbf{k}\right\}_{FS}}
\frac{\left|\mathbf{k}^{\prime}-\mathbf{k}\right|^{2}}{\omega_{\mathbf{\mathbf{k}^{\prime}-\mathbf{k}}}}
\delta\left(\omega-\omega_{\mathbf{\mathbf{\mathbf{k}^{\prime}-\mathbf{k}}}}\right)\nonumber\\
&\sim&  \int dQ\:\frac{Q^{3}}{\omega_{Q}}\delta\left(\omega-\omega_{Q}\right)
\sim  \omega^{2},
\label{Eq-Spectral-Function-Defect-free}
\end{eqnarray}
where we have made a change of variables, ${\bf Q}={\bf k}^\prime-{\bf k}$, and the sum over ${\bf k}$ 
was performed with the constraint that $\left|\mathbf{k}^{\prime}\right|=\left|\mathbf{k}\right|=k_{F}$:  
all states lie on the Fermi surface.

\subsection{Amorphous crystals: $\alpha^{2}\mathcal{F}_{\ell}\left(\omega\right)\sim\omega$}

Lattice distortion can be included into Eq.(\ref{Eq-Eliashberg-Spectral-Function}) 
by replacing $S_{\mathbf{q}}\left(\mathbf{k}^{\prime}-\mathbf{k}\right)$ with 
$\overline{S}^\ell_{\mathbf{q}}\left(\mathbf{k}^{\prime}-\mathbf{k}\right)$.
In the amorphous limit, $\ell\rightarrow a_\circ$, of the distorted structure factor in 
Eq. (\ref{Structure-Factor-Paracrystal}), we obtain, after subtracting the $\delta$-peak and 
retaining only terms with large ${\bf g}\neq 0$, 
$\lim_{\ell\rightarrow a_0}\overline{S}^\ell_{\mathbf{q}}\left(\mathbf{k}^{\prime}-\mathbf{k}\right)\rightarrow 1$. 
The interference pattern is now completely blurred and features concentric rings in reciprocal space.
Now the integrals over ${\bf Q}={\bf k}^\prime-{\bf k}$ and ${\bf q}$ become completely independent and
\begin{widetext}
\begin{equation}
\alpha^{2}\mathcal{F}_{\ell\rightarrow a_0}\left(\omega\right)  \sim
\sum_{\left\{\mathbf{k}^{\prime},\mathbf{k}\right\}_{FS},\mathbf{q},\nu}
\frac{\left|\mathbf{k}^{\prime}-\mathbf{k}\right|^{2}}{\omega_{\mathbf{\mathbf{q}},\nu}}
\cos^{2}\left(\hat{e}_{\nu}\left(\mathbf{q}\right);\widehat{\mathbf{k}^{\prime}-\mathbf{k}}\right)
\delta\left(\omega-\omega_{\mathbf{q},\nu}\right) 
\sim  \int dQ\:Q^{3}\:\left\langle 
\cos^{2}\left(\hat{e}_{\nu}\left(\mathbf{q}\right);\widehat{Q}\right)\right\rangle _{FS}
\int dq\:\frac{q^{2}}{\omega_{q}}\delta\left(\omega-\omega_{q}\right)
\sim  \omega.
\label{Eq-Spectral-Function-Amorphous}
\end{equation}
\end{widetext}
An immediate consequence is the transformation of the low frequency dependence of 
$\alpha^{2}\mathcal{F}_{\ell}\left(\omega\right)$ from $\sim\omega^2$, characteristic of 
clean metals as discussed previously, into $\sim\omega$, characteristic of amorphous metals \cite{Bergmann76-SC-AmorphousMetals-Review} . 
\subsection{Local lattice distortions: $\alpha^{2}\mathcal{F}_{\ell}\left(\omega\right)$ switching 
continuously from $\omega^2$ to $\omega$ behaviour}

For the intermediate distortion regime, where $a_0\ll\ell\ll\infty$, however, one observes a 
slowly but surely transfer of spectral weight from high to low-frequencies (the softening of 
the phonon spectrum), as can be seen from 
\begin{widetext}
\begin{equation}
\begin{multlined}
\alpha^{2}\mathcal{F}_{\ell}\left(\omega\right)\sim  
\sum_{\left\{\mathbf{k}^{\prime},\mathbf{k}\right\}_{FS},\mathbf{q},\nu}\overline{S}^\ell_{\mathbf{q}}\left(\mathbf{k}^{\prime}-
\mathbf{k}\right)\frac{\left|\mathbf{k}^{\prime}-\mathbf{k}\right|^{2}}{\omega_{\mathbf{\mathbf{q}},\nu}}
\cos^{2}\left(\hat{e}_{\nu}\left(\mathbf{q}\right);\widehat{\mathbf{k}^{\prime}-\mathbf{k}}\right)\delta\left(\omega-\omega_{\mathbf{q},\nu}\right)\\
\sim \int d^3{\bf q}\int_0^{2k_F} dQ\:Q^{3}\: \left\langle \cos^{2}\left(\hat{e}_{\nu}
\left(\mathbf{q}\right);\widehat{{\bf Q}}\right)\right\rangle _{FS}
\frac{\overline{S}^\ell_{\mathbf{q}}\left({\bf Q}\right)}{\omega_{q}}\delta\left(\omega-\omega_{q}\right),
\end{multlined} 
\end{equation}
\end{widetext}
retaining all allowed values for ${\bf g}$. The interference pattern is composed of a $\delta-$peak at ${\bf g}=0$,
a few clear peaks for small ${\bf g}$, but becomes ultimately blurred for larger ${\bf g}$. Now the integrals 
over $Q$ and $q$ are not independent, but convoluted by the electron-phonon structure factor. We can make 
use of the property (exact for $\ell\rightarrow\infty$ and approximate for $a_0\ll\ell<\infty$)
\begin{equation}
\overline{S}^\ell_{\mathbf{q}}\left({\bf Q}\right)\approx\frac{1}{Q^2}\frac{1}{\sin{\theta_{{\bf Q},{\bf q}}}}\overline{S}^\ell_{q}(Q)
\overline{S}^\ell_{\theta_{\bf q}}(\theta_{\bf Q})\overline{S}^\ell_{\varphi_{\bf q}}(\varphi_{\bf Q}),
\end{equation}
to perform the integration over $Q$, after which we end up with
\begin{equation}
\alpha^{2}\mathcal{F}_{\ell}\left(\omega\right)\sim \int d^3{\bf q}\:\frac{{\cal D}_{\ell}\left({\bf q},k_{F}\right)}{\omega_{q}}\delta\left(\omega-\omega_{q}\right),
\label{Eq-Spectral-Function-Defectal}
\end{equation}
where
\begin{widetext}
\begin{equation}
{\cal D}_{\ell}\left({\bf q},k_{F}\right)=\sum_{{\bf g}}\frac{1}{2\ell}
\left\{\ln{\left[\frac{1+\ell^2(|{\bf q}+{\bf g}|-2k_F)^2}{1+\ell^2|{\bf q}+{\bf g}|^2}\right]}
-2\ell |{\bf q}+{\bf g}|\left[\arctan{(\ell(|{\bf q}+{\bf g}|-2k_F))}
-\arctan{(\ell(|{\bf q}+{\bf g}|))}\right]\right\}.
\end{equation}
\end{widetext}
For the ${\bf g}=0$ term, this is a linear function of $q$, for $q\ll 2k_F$, in the $\ell\rightarrow\infty$ limit, 
thus reproducing the $\alpha^{2}\mathcal{F}_{\ell\rightarrow\infty}\left(\omega\right)\sim \omega^2$ result 
observed in simple metals. For ${\bf g}\neq 0$ and $|{\bf q}|\ll |{\bf g}|$, it reduces, in the extreme, amorphous, 
$\ell\rightarrow a_\circ$, limit, to a constant, thus giving rise to the $\alpha^{2}\mathcal{F}_{\ell\rightarrow\infty}\left(\omega\right)\sim \omega$ result 
observed in amorphous metals. Finally, for the intermediate distortion regime, $a_\circ\ll\ell<\infty$, 
it produces a slow transfer of spectral weight from high towards low frequencies, as expected 
for increasing lattice distortion, that can be simplified mathematically (after summation over the leading 
contributions of $|{\bf g}|\gg \frac{1}{a_\circ} $) as 
\begin{widetext}
\begin{equation}
    \alpha^{2}\mathcal{F}_\ell\left(\omega\right)
    \approx \left(1-\frac{2}{\pi k_F\ell}\right)\alpha^{2}\mathcal{F}_\infty\left(\omega\right)+
\frac{k_F^2 a_\circ^2}{12\, \ell}\alpha^{2}\mathcal{F}_{a_\circ}\left(\omega\right)
=\begin{cases}
               \alpha^{2}\mathcal{F}_\infty\left(\omega\right)\propto \omega^2             & \text{defectal-free } \ell \rightarrow \infty  \\
               \alpha^{2}\mathcal{F}_{a_\circ}\left(\omega\right)\propto \omega               & \text{amorphous } \ell\rightarrow a_\circ 
           \end{cases}
\label{Eq-Eliashberg-Spect-Fun-defectal-L}
\end{equation}
\end{widetext}
As is evident from the equation above, the spectral weight at low frequencies becomes
increasingly larger in a defectal-bearing system than in a pristine one. Such $\sim\omega$ 
behaviour is found in amorphous crystals \cite{Bergmann76-SC-AmorphousMetals-Review} 
and shows that one of the effects of defectals is to introduce damping to the phonon modes.

\end{document}